\documentclass[journal,twoside]{IEEEtran}
\usepackage{amsmath}
\usepackage{graphicx}
\usepackage{multirow}
\usepackage{amssymb}
\usepackage{array}
\usepackage[caption=false]{subfig}
\usepackage{cite}
\usepackage{color}
\usepackage{url}
\newcommand{\expect}{{\cal E}}

\newcommand{\PDCin}{P_{\text{in}}^{\text{DC}}}
\newcommand{\PDCPA}{P_{\text{PA}}^{\text{DC}}}
\newcommand{\PDCUSRP}{P_{\text{USRP}}^{\text{DC}}}
\newcommand{\PDCout}{P_{\text{out}}^{\text{DC}}}

\newcommand{\PRFin}{P_{\text{in}}^{\text{RF}}}
\newcommand{\PRFout}{P_{\text{out}}^{\text{RF}}}
\newcommand{\etaDCtoDC}{\eta_{\text{DC-to-DC}}}
\newcommand{\etaDCtoRF}{\eta_{\text{DC-to-RF}}}
\newcommand{\etaRFtoRF}{\eta_{\text{RF-to-RF}}}
\newcommand{\etaRFtoDC}{\eta_{\text{RF-to-DC}}}
\newcommand{\fc}{f_{\rm c}}
\newcommand{\xB}{x_{B}}
\newcommand{\xI}{x_{\rm I}}
\newcommand{\xQ}{x_{\rm Q}}

\newcommand{\xbar}{\bar{x}}
\newcommand{\xhat}{\hat{x}}
\newcommand{\xBbar}{\bar{x}_{B}}
\newcommand{\xIbar}{\bar{x}_{\rm I}}
\newcommand{\xQbar}{\bar{x}_{\rm Q}}
\newcommand{\pai}{\rm{PA}_{\rm I}}
\newcommand{\pae}{\rm{PA}_{\rm E}}
\DeclareMathOperator{\Real}{Re}

\usepackage[left,pagewise,switch]{lineno}

\begin{document}

\title{Waveforms and End-to-End Efficiency\\in RF Wireless Power Transfer\\Using Digital Radio Transmitter}

\author{Nachiket Ayir,~\IEEEmembership{Student Member,~IEEE,}
Taneli Riihonen,~\IEEEmembership{Member,~IEEE,}\\
Markus All{\'e}n,
and~Marcelo Fabi{\'a}n Trujillo Fierro%
\thanks{Manuscript received November 16, 2019; revised June 18, 2020 and September 11, 2020; accepted November 30, 2020. Date of publication Month ??, 2021; date of current version \today. This work was partially supported by the Ulla Tuominen foundation and the Academy of Finland under the grants 310991/326448 and 315858. {\it (Corresponding author: Nachiket Ayir).}}%
\thanks{The authors are with the Faculty of Information Technology and Communication Sciences, Tampere University, Finland (e-mail: \{nachiket.ayir, taneli.riihonen, markus.allen\}@tuni.fi, marcelotrujillo83@gmail.com).}% 
\thanks{This paper is an expanded version from the IEEE MTT-S International Microwave Symposium (IMS 2019), Boston, MA, USA, June 2-7, 2019. \cite{Nachiket-IMS}}% 
\thanks{Color versions of one or more of the figures in this article are available
online at \protect\url{https://ieeexplore.ieee.org}}% 
\thanks{Digital Object Identifier 10.1109/TMTT.XXXX.XXXXXXX}% 
}

\markboth{IEEE Transactions on Microwave Theory and Techniques,~Vol.~XX, No.~XX, Month~2021}%
{Ayir \MakeLowercase{\textit{et al.}}: Waveforms and End-to-End Efficiency in RF Wireless Power Transfer Using Digital Radio Transmitter}

\maketitle

%%%%%%%%%%%%%%%%%%%%%%%%%%%%%%%%%%%%%%%%%%%%%%%%%%%%%%%%%%%%%%%%%%%%%%%%%%%%%%%%%%%%%%%%%%%%%%%%%%%%%%%%%%%%%%%%%%%%%%%%%%%%%%%%%%%%%%%%%%%%%%%%%%%%%%%%%%%%%%%%
\begin{abstract}
We study radio-frequency (RF) wireless power transfer (WPT) using a digital radio transmitter for applications where alternative analog transmit circuits are impractical. An important parameter for assessing the viability of an RF WPT system is its end-to-end efficiency. In this regard, we present a prototype test-bed comprising a software-defined radio (SDR) transmitter and an energy harvesting receiver with a low resistive load; employing an SDR makes our research meaningful for simultaneous wireless information and power transfer (SWIPT). We analyze the effect of clipping and non-linear amplification at the SDR on multisine waveforms. Our experiments suggest that when the DC input power at the transmitter is constant, high peak-to-average power ratio (PAPR) multisines are unsuitable for RF WPT over a flat-fading channel, due to their low average radiated power. The results indicate that the end-to-end efficiency is positively correlated to the average RF power of the waveform, and that it reduces with increasing PAPR. Consequently, digital modulations such as phase-shift keying (PSK) and quadrature amplitude modulation (QAM) yield better end-to-end efficiency than multisines. Moreover, the end-to-end efficiency of PSK and QAM signals is invariant of the transmission bit rate. An in-depth analysis of the end-to-end efficiency of WPT reveals that the transmitter efficiency is lower than the receiver efficiency. Furthermore, we study the impact of a reflecting surface on the end-to-end efﬁciency of WPT, and assess the transmission quality of the information signals by evaluating their error vector magnitude (EVM) for SWIPT. Overall, the experimental observations of end-to-end efficiency and EVM suggest that, while employing an SDR transmitter with fixed DC input power, a baseband quadrature PSK signal is most suitable for SWIPT at large, among PSK and QAM signals.
\end{abstract}


%%%%%%%%%%%%%%%%%%%%%%%%%%%%%%%%%%%%%%%%%%%%%%%%%%%%%%%%%%%%%%%%%%%%%%%%%%%%%%%%%%%%%%%%%%%%%%%%%%%%%%%%%%%%%%%%%%%%%%%%%%%%%%%%%%%%%%%%%%%%%%%%%%%%%%%%%%%%%%%%
\section{Introduction}

\IEEEPARstart{T}{he} fifth generation (5G) of mobile communications is expected to network trillions of sensors into the so-called Internet-of-Things (IoT); such an extreme-scale deployment would certainly require equipping the sensors with replenishable energy sources. Perhaps an even bigger concern would be the looming environmental hazard from the failure to recycle IoT-connected disposables, if they comprised ordinary batteries with toxic materials. In recent years, far-field radio-frequency (RF) {\em wireless power transfer} (WPT) is being studied as a possible solution to these challenges~\cite{1G_Mobile_Power_Networks}. It would allow powering sensors on demand, and supercapacitors could suffice for temporary energy storage, thus averting sensors from becoming hazardous waste after their working life.

\subsection{Motivation}

Commercial products for short-range (about an inch) and mid-range (about $1-2$~m) WPT are already available. While the former is solely based  on magnetic induction coupling, the latter works on magnetic resonance-based inductive coupling \cite{Magnetic_resonance_Kurs}, which relies on resonating a secondary/primary coil for WPT. The former comes  with its own specifications \cite{Inductive_WPT_specs} and a separate Qi standard \cite{Qi_standard} adopted by the Wireless Power Consortium, while the latter is regulated by the AirFuel alliance \cite{AirFuel}. Both these competing technologies have their own benefits and limitations \cite{A_review_of_WPT,comparison_induction_resonance_WPC,comparison_induction_resonance_DigiKey,comparison_induction_resonance_RFWPT_Ansys}. However, these are devoid of providing mobility or long-range operation. Far-field RF WPT has the potential to overcome these issues.

\begin{figure*}[t]
	\centering
	\subfloat[general block diagram]{\includegraphics[width=0.45\textwidth]{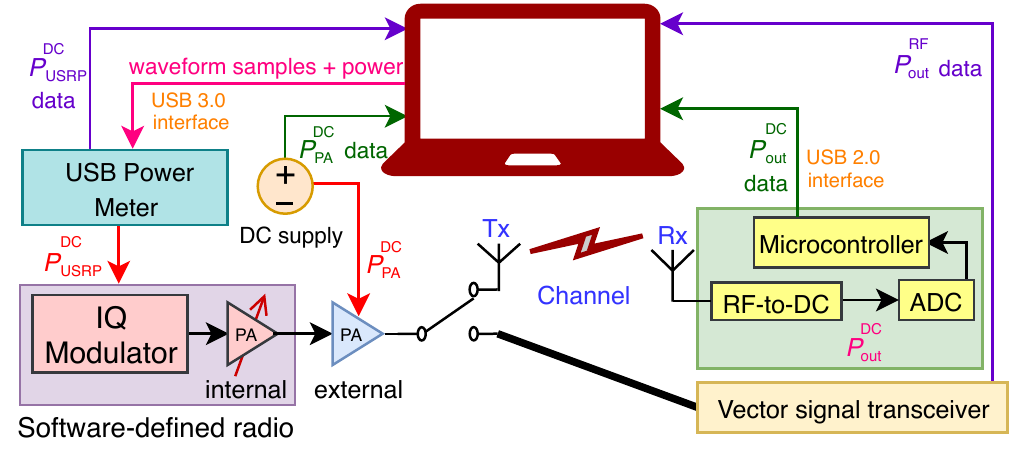}\label{fig:sys_a}}
	\hspace{1pt}
	\subfloat[measured average transmit RF power]{\includegraphics[width=0.265\textwidth]{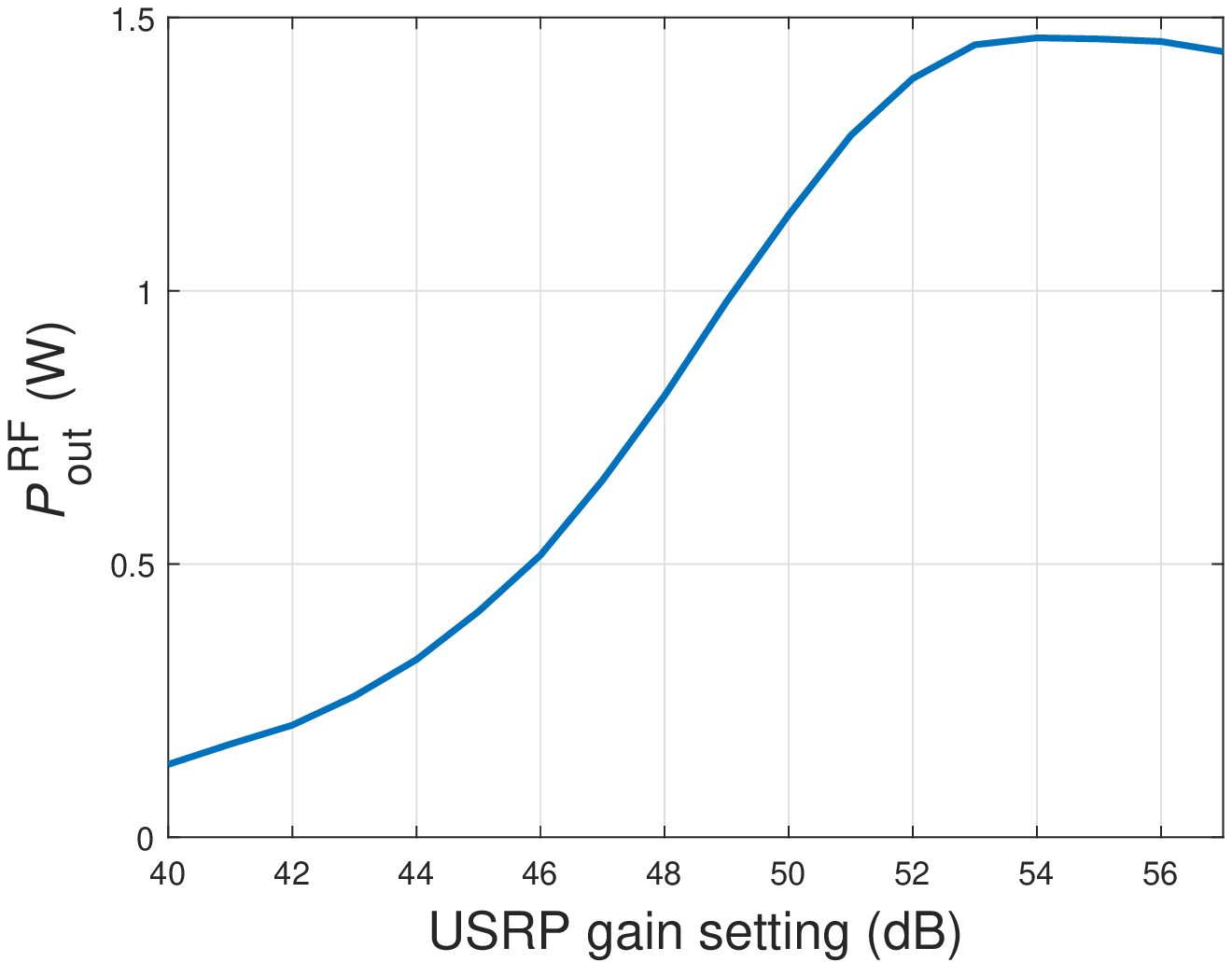}\label{fig:sys_b}}\hspace{1mm}
	\subfloat[measured average harvested DC power]{\includegraphics[width=0.265\textwidth]{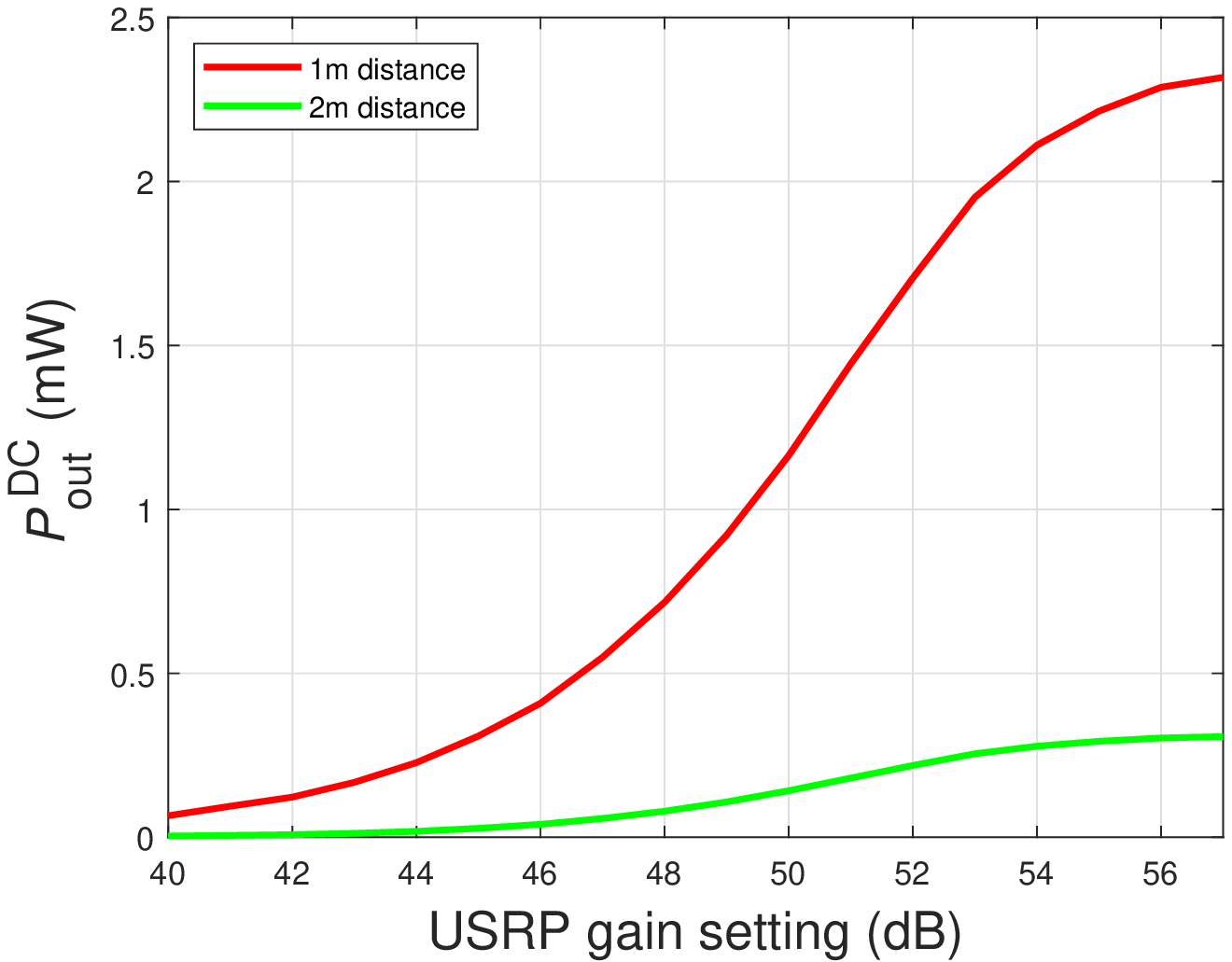}\label{fig:sys_c}}
	\caption{\label{fig:sys}Block diagram of the presented RF WPT system along with an example of measured average transmit RF power ($\PRFout$) and the corresponding average harvested DC power ($\PDCout$) for two transmitter--receiver distance values, while varying the gain setting of the USRP. The baseband signal is a $200$~kHz complex tone with unit amplitude. The total DC power consumption at the transmitter ($\PDCin$) decreases marginally from $12.4$~W to $11.4$~W as the USRP gain setting is increased from $40$~dB to $57$~dB. The decrease in $\PRFout$ can be attributed to the saturation of the external PA at higher USRP gain settings.}
\end{figure*}
It has conventionally been believed that RF WPT concepts would be unable to deliver sufficient energy to sensors for their operation, due to the undeniably huge over-the-air propagation losses. However, the energy consumption of simple devices for sensing and computing has fallen drastically over the years in accordance to the Koomey's law~\cite{Koomey}, which has reignited the research interest in RF WPT of late. Furthermore, the wireless communications research community has been keen on developing and harnessing {\em simultaneous wireless information and power transfer} (SWIPT)~\cite{SWIPT-Krikidis}, which allows convenient reuse of spectrum and energy, since many sensors are targets for both transmissions anyway.

A central research theme in the evolution of WPT and SWIPT systems is their power efficiency~\cite{1G_Mobile_Power_Networks}. This is because the true technical challenge in WPT is to reduce the required transmission power and overall energy losses enough so that the technology could be considered economically viable, sustainable and safe. Likewise, in this work, we study the end-to-end WPT efficiency that is the ratio of the direct current (DC) power harvested at a receiver ($\PDCout$) to all the DC power supplied to a transmitter ($\PDCin$):

\begin{equation}
\label{eq:etaDCtoDC} 
    \etaDCtoDC = \frac{\PDCout}{\PDCin} = \etaDCtoRF \cdot \etaRFtoRF \cdot \etaRFtoDC,
\end{equation}
which is further split into the RF waveform generation efficiency at the source transmitter ($\etaDCtoRF$), the overall RF transmission efficiency between antenna connectors ($\etaRFtoRF$), and the power conversion efficiency at the destination {\em energy harvesting} (EH) receiver ($\etaRFtoDC$).

\subsection{Literature Review}
The large body of recent academic research on WPT, SWIPT, and RF EH has already developed solid theoretical foundations for many key aspects such as rectifier design, waveform design, beamforming algorithms, link-level analysis, and optimization. Comprehensive surveys of such research works are readily available in~\cite{reference,Survey-Hu,Survey-Tharindu,Survey_beamforming,Survey_efficiency,Survey_networks}. A major part of the current literature focuses mainly on optimizing $\etaRFtoDC$. For instance, the authors in \cite{Magazine_Carvalho} experimentally deduce that multisine signals are optimal for maximizing $\etaRFtoDC$ in comparison to orthogonal frequency-division multiplexing (OFDM) signals, chaotic signals, harmonic signals, etc.

Moreover, there are major theoretical contributions on rectifier modeling, relying on approximations of the Shockley diode equation \cite{BC1,Carvalho_diode_equation} or data-based curve-fitting \cite{Elena1} to mathematically characterize the non-linearity of diodes. The idea of rectifier modeling has been further extended to address problems like waveform design under the availability of complete \cite{BC1,Rui1} or limited \cite{BC2,Kim_imperfectCSI} knowledge of the channel state information (CSI).

In the complete absence of CSI, the impact of fading and transmit diversity on signal design for WPT has been studied in \cite{Clerckx_transmit_diversity}. Additionally, in the case of imperfect CSI, the problem of simultaneously powering multiple non-linear rectifier-based sensors with RF power by employing robust beamforming at the transmitter was studied in \cite{Elena_Robust_Beamforming,Sun_Robust_Beamforming}.

The research on communication and signal design for optimizing the  end-to-end transmission in RF WPT is already a flourishing area with many noteworthy contributions  \cite{BC1,bruno_arxiv,bruno_Communications_and_Signals_Design_for_WPT}, assiduously summarized in \cite{reference}. Besides mathematical modeling,  the physical design of a rectenna (a portmanteau for the combination of a receiver antenna and an RF diode-based rectifier) operating at $2.4$~GHz, by designing Schottky diode-based voltage rectifiers, is presented in \cite{Clerckx_rectenna, Rectenna_collado}.

The researchers in \cite{reference, Magazine_Carvalho, BC1, bruno_arxiv} have reported that a high peak-to-average power ratio (PAPR)  co-phased multisine waveform is suitable for transmission over a flat-fading channel. Most of these studies employ power oscillators or scientific instruments like vector signal transceivers (VSTs) or arbitrary waveform generators as transmitters to generate high-PAPR multisine signals. Besides these, switched-mode power amplifiers, such as current-mode class D converters, could also generate high-PAPR signals in a single-stage transmitter. 

Furthermore, to evade the challenges associated with high-PAPR signal generation in a single transmitter,  spatial combining techniques have been explored \cite{Adib_Spatial, Carvalho_spatial_combining}. While \cite{Adib_Spatial} determines the phases for multiple distinct sinusoids  transmitted by an external source, \cite{Carvalho_spatial_combining} employs multiple synchronized transmitters to generate a high-PAPR signal at the receiver. 

The designs in \cite{reference, Magazine_Carvalho, BC1} that lead to the claim of high-PAPR signals being suitable for RF WPT over a flat-fading channel do not demonstrate experimentally the effect of the presence of a non-linear power amplifier (PA) at the transmitter. Though the experimental setup and the measurement results in \cite{bruno_arxiv} capture any potential effect of the PA non-linearity, the effect of the PA non-linearity and of $\etaDCtoRF$ are not explicitly studied and highlighted.

Ideally, efficient diode-based EH would require the received signals to have as high PAPR as possible (with an impulse train being the ultimate waveform), while in applications such as SWIPT, employing alternative RF signal generation and amplification architectures is challenging in the first place.  Moreover, the problem with maximizing PAPR is that it disregards the limited output range of the digital-to-analog converter (DAC) of any digital transmitter, as well as the non-linearity which arises as the amplitude peaks of a high-PAPR waveform drive the PA into saturation, thereby distorting the shape of the actual transmitted waveform. With their envelope distorted, the multisine waveforms might not remain optimal for WPT anymore. In literature, there is thus clearly a dearth of research studies that explain the impact of using a practical software-defined radio (SDR) transmitter for RF WPT.

In literature, we encounter several studies that have developed WPT \cite{GKKurt,Kurt_Effects_of_Different_Modulation_Techniques_on_Charging_Time_in_RF_Energy_Harvesting_System,Tsolovos,SJha1,GChen,Rabaey,RHoward,Clerckx_rectenna} and SWIPT \cite{SWIPT_Clerckx,SWIPT_Claessens} test-beds for varying purposes, from simply charging a battery to studying the effects of the factors influencing system performance. However, this research direction is still fairly in its nascent stage and provides opportunity for original contributions.

\subsection{Contributions and Organization}
All the aforementioned test-beds operate in the $915$~MHz or $2.4$~GHz industrial, scientific, and medical (ISM) bands using a Universal Software Radio Peripheral (USRP) or the rather expensive VST hardware. For our study, we employ a USRP while operating in the $863-873$~MHz European unlicensed band\footnote{The conclusions of this work are applicable to other ultra-high frequency (UHF) bands as well, whenever bandwidth is not large. Some related results for the American $902-928$~MHz ISM band have been presented in \cite{Nachiket-IMS}.}. The block diagram of our test-bed along with a few preliminary results is presented in Fig.~\ref{fig:sys}. These would be explained in detail in Sections~\ref{sec:test-bed} and~\ref{sec:experiments}, respectively.

The novel contributions of this paper are as follows.
\begin{enumerate}
\item
We present an original GNU Radio-based prototype test-bed for end-to-end (i.e., DC-to-DC) RF WPT comprising a USRP transmitter and an energy harvesting receiver with a diode-based rectifier as illustrated in Fig.~\ref{fig:sys}.  Our test-bed allows us to study $\etaDCtoDC$ and $\etaDCtoRF$, while investigating the use of a digital radio for transmission.

\item
We show that with a fixed $\PDCin$ and the resistive loads in the orders of a few hundred ohms, ordinary communication signals such as PSK and QAM are more suitable for RF WPT than the co-phased multisine waveforms with high PAPR that are tailored for EH only. Under the same $\PDCin$ constraint, when the transmitter is a digital IQ modulator (which is the case with an SDR), we observe that the high-PAPR multisines get either clipped or severely distorted by the digital radio transmitter, thus diminishing their efficacy in RF WPT. 

\item
Our experimental results demonstrate the novel fact that typically receiver efficiency is better than transmitter efficiency ($\etaRFtoDC  > \etaDCtoRF$). Moreover, while the latter remains roughly constant for given amplifier gain setting so that it can be improved only by transmitter design, the former can be enhanced during operation by delivering more power to the energy harvesting receiver through beamforming or reducing propagation distance.

\item
We evaluate the effect of imperfections in an SDR transmitter on the quality of RF signal transmission in terms of error vector magnitude (EVM). The study reveals that DAC clipping deteriorates the EVM of multisine signals as their amplitude increases. On the other hand, the EVM for communication signals turns out to be fairly invariant of their bit rate. 

\end{enumerate}

With respect to the first point above, the other known test-beds for RF WPT/SWIPT such as  \cite{Clerckx_rectenna, SWIPT_Clerckx} (and also \cite{bruno_arxiv} partly) employ expensive scientific instruments, such as National Instruments (NI) FlexRIO (PXI-7966R) FPGA module and transceiver adapter module (NI 5791R), as transmitters. Contrarily, a USRP is a de-facto standard prototyping device for wireless test-beds as it is both low-cost and represents the limitations of a practical transmitter in a real-world setup.

It should be noted that the observation in the second point above is in contrast with that in \cite{bruno_arxiv}, where the researchers show that CSI-adaptive multisine waveforms outperform PSK and QAM in terms of the energy harvested. The reasons for this disparity are explained in Section~\ref{sec:experiments} in detail, but briefly stated they stem in general from the differences in receiver-side equipment and resistive loads in the two research setups.

The remainder of this article is organized as follows. In the next section, we study the various impairments, which the multisine signals suffer from before transmission from a digital radio transmitter, through theoretical analysis and simulations. Next, we present the details of the test-beds employed in this study in Section~\ref{sec:test-bed}. We conducted experiments employing our test-beds to determine $\etaDCtoDC, \etaDCtoRF, \etaRFtoRF$ and EVM for different test waveforms. The details of these experiments and their key observations are reported in Section~\ref{sec:experiments}. Finally, we present the conclusions of this work in Section~\ref{sec:conclusions}.

%%%%%%%%%%%%%%%%%%%%%%%%%%%%%%%%%%%%%%%%%%%%%%%%%%%%%%%%%%%%%%%%%%%%%%%%%%%%%%%%%%%%%%%%%%%%%%%%%%%%%%%%%%%%%%%%%%%%%%%%%%%%%%%%%%%%%%%%%%%%%%%%%%%%%%%%%%%%%%%%
\section{Digital Transmission of Multisine signals}\label{sec:DAC}
Let us analyze the detrimental effects of the practical limitations of digital transmission on multisine signals following stage-by-stage the common IQ modulator architecture. 

Without loss of generality, we assume that the digital baseband signal to be transmitted is a complex-valued $N$-tone multisine waveform given by
\begin{align}\label{multisine}
    x(t) &= \xI(t) + j\,\xQ(t) \nonumber \\
         &= \sum_{n=1}^{N} A_{n} \exp\left(j\,2\pi f_{n}\,t + j\,\phi_{n}\right),
\end{align}
where $A_n$, $f_{n}$ and $\phi_{n}$ represent the amplitude, the baseband frequency and the phase of the $n$th tone, respectively. A multisine waveform can approximate (or, if $N\to\infty$, exactly represent) any periodic signal $x(t)$, when interpreted as a Fourier series, which justifies the assumption.

The process of generation of an analog RF signal from discrete IQ samples is well known, so we avoid explaining it here. While generating an RF waveform from discrete multisine samples, a digital radio introduces non-linear distortions in the signal. These distortions would alter the shape (and hence the PAPR) of the modulated RF waveform owing to which it may not remain suitable for RF WPT over a flat-fading channel.  The sources of the distortions, in a digital radio, include the DAC and amplification by the internal PA ($\pai$). The former introduces distortion due to clipping effect and quantization, while the latter introduces distortion because of non-linear input--output transfer characteristics. Let us study these through mathematical analysis and simulations.

\subsection{Clipping and Limited Resolution in Digital-to-Analog Conversion}
Any DAC employed in digital radios has inherently some limited output range for discrete IQ samples; we set it to ${\pm} 1$ herein without loss of generality.  Consequently, any sample of $\xI(t)$ or $\xQ(t)$ with magnitude greater than one is truncated to one. The resultant analog baseband waveform generated from the digital baseband samples is of the form 
\begin{equation}
\xbar(t) = \xIbar(t) + j\,\xQbar(t)
\end{equation}
for which in each branch ($B \in \{{\rm I}, {\rm Q}\}$)
\begin{equation}
    \xBbar(t)=
    \begin{cases}
        +1,   & \xB(t) \ge 1,\\    
        -1,   & \xB(t) \le -1,\\
        \xB(t), & \text{otherwise.}
    \end{cases}
    \label{eq:clipping}
\end{equation}
The bar signifies the clipping of the original waveform which would generate non-linear distortion. The resultant RF waveform is given by
\begin{align}\label{rf}
\xhat(t) = \sqrt{2} \, \Real\left\{\xbar(t) \cdot \exp\left(j\,2\pi \fc\,t\right)\right\},
\end{align}
where $\Real\{.\}$ represents the real part of a complex number and $\fc$ denotes the carrier frequency of the transmission. 

Whether the analog baseband waveform would get clipped or not depends on the amplitudes of the individual baseband sinusoids ($A_n$), for which we can find two cases as follows.

\subsubsection{Case I}
If $A_n$, $n=1, 2, \ldots, N$, are low enough to prevent clipping, then the shape and, hence, the PAPR of the waveform remain intact. In this case, the resultant RF waveform before power amplification is given by
\begin{align}
\xhat(t) = \sqrt{2} \, \sum_{n=1}^{N} A_{n} \cos\left( 2\pi (\fc + f_{n})\,t + \phi_{n} \right)
\end{align}
with a peak power of 
\begin{equation}\label{xrfpeakcase1}
    \{\xhat^2(t)\}_{\rm peak}
        \begin{cases}
        =2~( \sum_{n=1}^{N} A_{n})^2,  & \text{when for } n=1,2,\ldots, N,\\ &\hspace{-40pt}\phi_n \equiv -2\pi (\fc + f_{n})\,t_{\rm peak} \text{ mod } 2\pi,\\    
        <2~( \sum_{n=1}^{N} A_{n})^2,   & \text{otherwise,}
    \end{cases}
\end{equation}
where $\{.\}_{\rm peak}$ represents the peak value of a real signal and $t_{\rm peak}$ is the peak time instant at which all the phases align. The  corresponding average power is given by
\begin{equation}\label{avgpower}
\expect\{\xhat^2(t)\} = \expect\{|x(t)|^2\} = \sum_{n=1}^{N} A_{n}^2.
\end{equation}

We observe that the average power of the RF waveform depends on the amplitudes of the individual tones. Thus, lowering $A_{n}$ to avoid clipping and preserve the PAPR will lower the average radiated RF power. The PAPR of the RF waveform in this case becomes 
 \begin{equation}\label{PAPR1}
     {\rm PAPR} \leq \frac{2~(\sum_{n=1}^{N} A_{n})^2}{\sum_{n=1}^{N} A_{n}^2}
 \end{equation}
and depends on the values of $\phi_{n}$ as specified in \eqref{xrfpeakcase1}. 

\subsubsection{Case II}
If the values of $A_n$ are such that the resultant amplitude of $\xB(t)$ is greater than one, then the DAC limits the amplitude of $\xB(t)$ to one. In this case, the average power and, consequently, the PAPR of the RF waveform $\xhat(t)$ can be computed by piecewise integration.

We consider the following assumptions\footnote{The assumptions made in Case II stem from our observation that the channel response is essentially frequency-flat in the bandwidth of interest ($868-872$~MHz). This is shown in Fig.~\ref{fig:channel_response}. For a flat-fading channel, uniform power allocation across the frequencies of a multisine signal is recommended [13]. Hence, we can keep the magnitudes and phases the same across frequencies to attain a high-PAPR signal.} to simplify the analysis: $A_{n}=A, f_{n}=n f_{0}, \phi_{n}=\phi$. With these assumptions, we can express the co-phased $N$-tone multisine in \eqref{multisine} as 
\begin{align}
     x(t)
     &= A \cdot \exp\left(j\,\phi\right) \sum_{n=1}^{N} \exp\left(j\,2\pi n f_{0} t \right), %\\
\end{align}
where the in/quadrature-phase (I/Q) components are given as
\begin{align}\label{xi}
    \xI(t) &= A \cdot \frac{\sin\left(N \pi f_{0} t\right) \cdot \cos\left((N+1)\pi f_{0} t + \phi\right)}{\sin\left( \pi f_{0} t\right)}, \\
    \xQ(t) &= A \cdot \frac{\sin\left(N \pi f_{0} t\right) \cdot \sin\left((N+1)\pi f_{0} t + \phi\right)}{\sin\left( \pi f_{0} t\right)}.
\end{align}

It is evident from \eqref{xi} that the exact time instants at which $\xI(t)$ gets clipped can only be found by numerical analysis, which is as follows. Without any loss of generality, consider a single period of the multisine waveform $\xI(t)$ from $0 < t \leq T$, where $T=1/f_0$. Let $t_{1},t_{2}, \ldots t_{M}$ be the time instants between which $\xI(t)$ is clipped and not clipped by the DAC, such that $0 < t_1 < t_2 < \cdots < t_M < T$. Thus, we have
\begin{equation}\label{xi1}
|\xIbar(t)| = 
\begin{cases}
        1,        & t \in [0,t_1],[t_2,t_3],\ldots,[t_M,T],\\
        |\xI(t)|, & t \in (t_1,t_2),(t_3,t_4),\ldots,(t_{M-1},t_M).
    \end{cases}
\end{equation}
From \eqref{xi1}, it is clear that the peak power of $\xIbar(t)$ would be one and that its average power can be computed as
\begin{align}\label{xiavg}
    \expect\{\xIbar^2(t)\}
   & =  \frac{1}{T} \Big[ t_1 + (t_3 - t_2) + \ldots + (T - t_M) + \\
      &\int_{t_1}^{t_2} \xI^2(t) dt + \int_{t_3}^{t_4} \xI^2(t) dt + \ldots + \int_{t_{M-1}}^{t_M} \xI^2(t) dt \Big].
      \nonumber 
\end{align}
The peak power and average power for $\xQbar(t)$ can be computed on similar lines and are, thus, omitted here. The total average power of a complex signal $\xbar(t)$ is given by
\begin{equation}\label{clipped_bb_avg}
\expect\{|\xbar(t)|^2\} =  \expect\{\xIbar^2(t)\} + \expect\{\xQbar^2(t)\}.
\end{equation}

\begin{figure*}[t]
	\centering
	\subfloat[]{\includegraphics[width=0.3\textwidth,trim={20 3 35 0}]{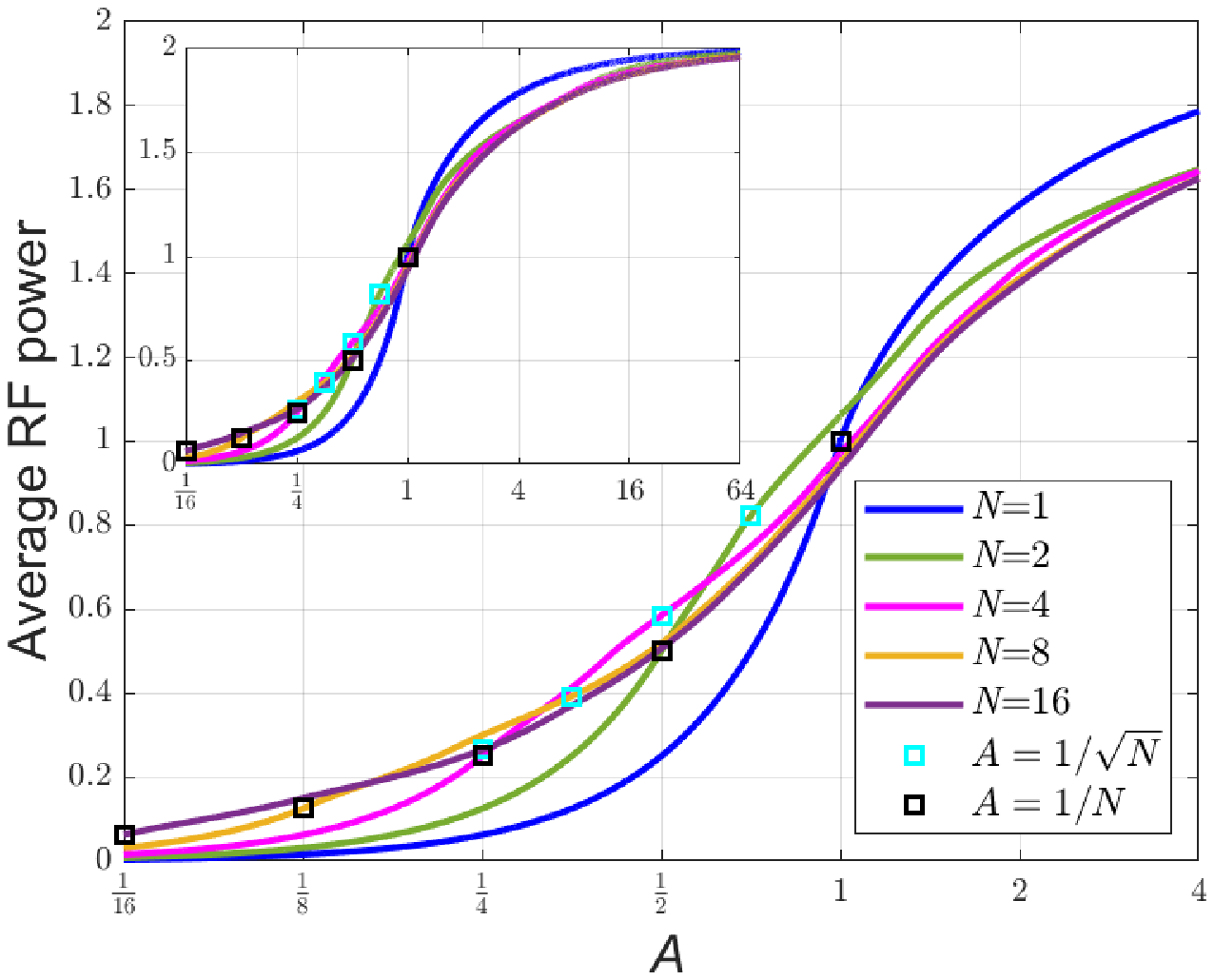}\label{fig:DAC_efficiency_a}} \hspace{5pt}
	\subfloat[]{\includegraphics[,width=0.3\textwidth]{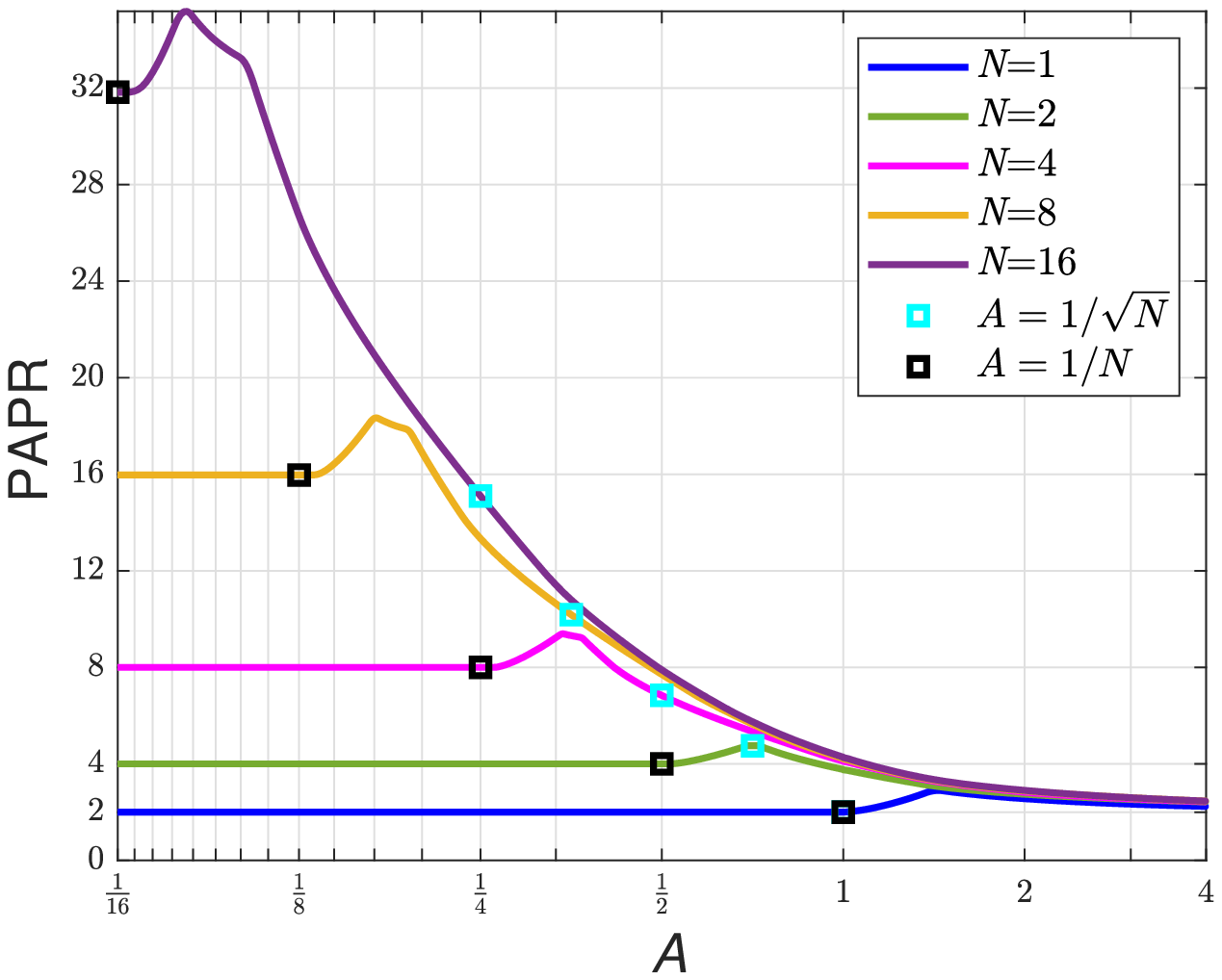}\label{fig:DAC_efficiency_b}}
	\hspace{5pt}
    \subfloat[]{\includegraphics[width=0.3\textwidth]{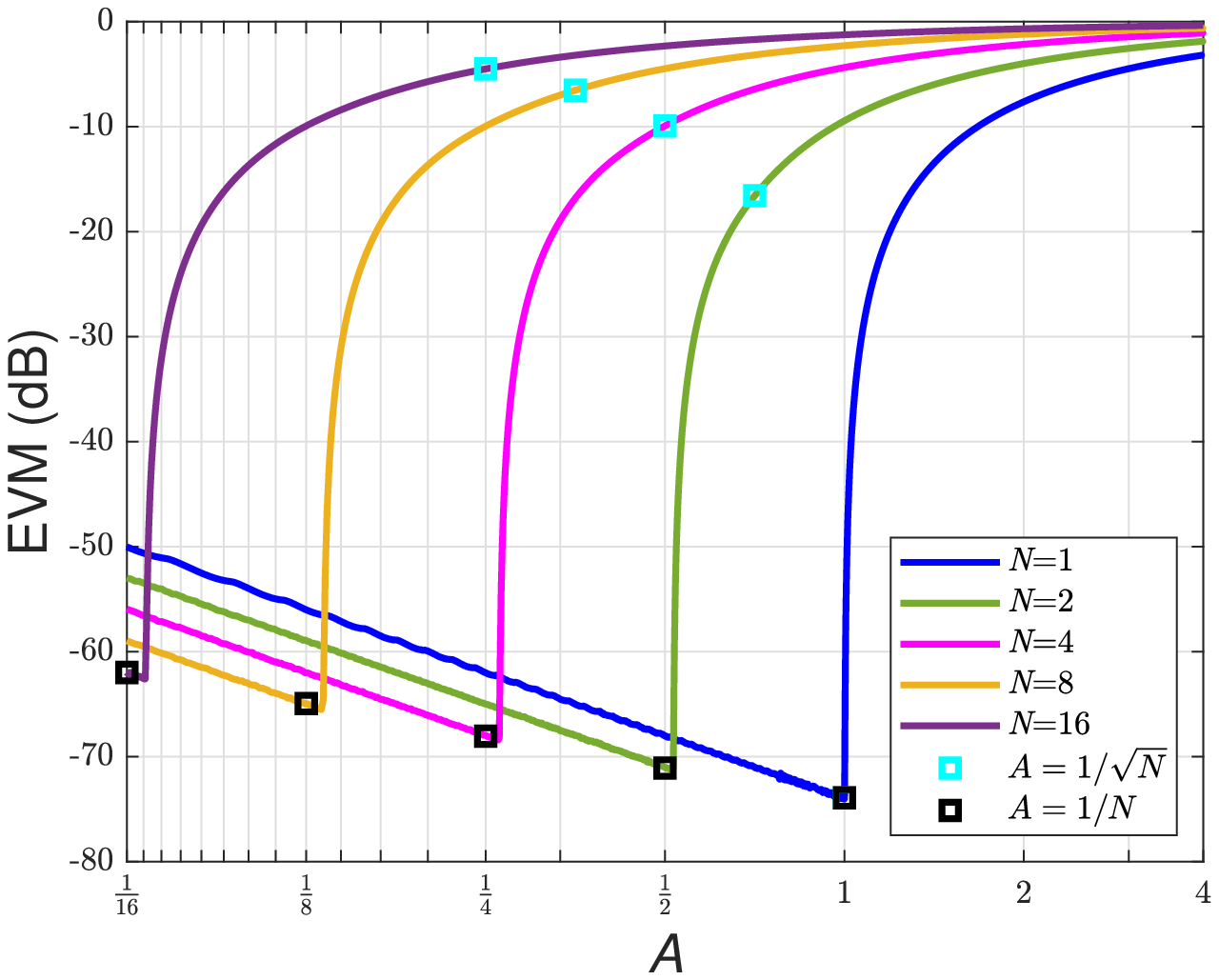}\label{fig:DAC_efficiency_c}}
	\caption{\label{fig:DAC_efficiency}Simulation-based plots showcasing the impact of varying the amplitude $A$ of $N$-tone multisines on the (a) average power $\expect\{\xhat^2(t)\}$, (b) PAPR and (c) EVM of the RF waveform (before $\pai$), for different values of $N$.}
\end{figure*}

On simplifying \eqref{rf} and considering a clipped baseband multisine waveform, the RF signal can be expressed as
 \begin{equation}\label{rf2}
     \xhat(t) = \sqrt{2} \cdot \Big(\cos(2 \pi f_c t) \cdot \xIbar(t) - \sin(2 \pi f_c t) \cdot \xQbar(t) \Big),
 \end{equation}
and the corresponding average power is given as 
\begin{subequations}
\begin{align}
\expect\{\xhat^2(t)\} 
&= \,\expect\{\xbar^2(t)\} - \expect\{2\cdot\sin(4 \pi f_c t) \cdot \xIbar(t)\,\xQbar(t)\} \nonumber\\
&\quad + \expect\{\cos(4 \pi f_c t)\cdot(\xIbar^2(t)-\xQbar^2(t))\},\label{clippedrfavg_a}
\end{align}
which becomes
\begin{align}
\expect\{\xhat^2(t)\}&\begin{cases}
        = \expect\{\xbar^2(t)\},   & \phi = \frac{\pi}{4}, \\
        \approx \expect\{\xbar^2(t)\},  & \textrm{otherwise.} 
    \end{cases}\label{clippedrfavg_b}
\end{align}
\end{subequations}

We observe that the average power of the RF waveform is approximately equal to the average power of the clipped baseband waveform in general case. It is exactly equal when the third term in \eqref{clippedrfavg_a} would integrate to zero. We found that this case arises when $\phi = \frac{\pi}{4}$, which yields $\xI(t) = \xQ(T-t)$.

Due to clipping, the RF waveform would achieve its peak power at time instants  when both $|\xIbar(t)| = |\xQbar(t)| = 1.$ Substituting these in \eqref{rf2}, we observe that during those instants the RF waveform will have a peak power of four. Thus, for the case of a clipped multisine, the PAPR of the RF waveform can be computed as
 \begin{equation}\label{PAPR2}
     {\rm PAPR} = \frac{4}{\expect\{\xhat^2(t)\}}
     \approx
     \frac{4}{\expect\{\xbar^2(t)\}}.
 \end{equation}

The second source of distortion is the limited resolution of the DAC which results in quantization errors that may creep in if the fast varying tails of the multisine signals lie in between any two quantization levels.

\subsection{Simulations and Observations}
Let us observe the impact of both the aforementioned effects on the PAPR, the average RF power, and the error vector magnitude (EVM) of multisine waveforms before amplification by the $\pai$, through MATLAB simulations. For the simulations, we set $A_n = A$ and $\phi_n = \frac{\pi}{4}$.\footnote{The optimal way to obtain an $N$-tone multisine in a digital radio is by setting $\phi_n = \frac{\pi}{4}$, which results in $\xI(t) = \xQ(T-t)$. Moreover, the wireless channel is essentially almost flat for the signal bandwidth under consideration (cf.\ Section~\ref{subsection:channel_sounding}). So, we choose a common phase for all the harmonics to obtain a co-phased multisine signal, as recommended in \cite{BC1,bruno_arxiv,bruno_Communications_and_Signals_Design_for_WPT}.} The results are as shown in Fig.~\ref{fig:DAC_efficiency}. In the subfigures therein, we have plotted the responses of multisines with different $N \in \{1,2,4,8,16\}$,  for varying $A$. As reference markers, we have used two cases: $A = \frac{1}{N}$ and $A = \frac{1}{\sqrt{N}}$. The former ensures that $\xB(t) < 1$, thus evading any clipping by the DAC, while the latter assures that the digital samples of all the multisine signals will have the same average power. This allows for a fair performance comparison.

The observations, and their explanations, are as follows.
\begin{enumerate}
\item Fig.~\ref{fig:DAC_efficiency_a} shows the variation of the average RF power. We observe that  $\expect\{\xhat^2(t)\}$ does not increase linearly with $A$. The slope of the curve increases up to $A = \frac{1}{N}$, and then decreases. The reduction in the rate of increase of $\expect\{\xbar^2(t)\}$ is due to the onset of DAC clipping after $A = \frac{1}{N}$. The inset figure depicts that increasing $A$ further eventually saturates $\expect\{\xbar^2(t)\}$ to two. This occurs since the clipping is so severe that we are left with a waveform similar to a square wave at baseband irrespective of $N$. Considering multisine signals with same input average power $(A = \frac{1}{\sqrt{N}})$, we observe that $\expect\{\xhat^2(t)\}$ decreases with increasing $N$, with a single sinusoid  baseband signal achieving the highest $\expect\{\xhat^2(t)\}$.

\item The variation of PAPR of the RF waveform w.r.t.\ $A$ is shown in Fig.~\ref{fig:DAC_efficiency_b}. While $A\leq \frac{1}{N}$, the multisines are not clipped and, thus, ${\rm PAPR} = 2 N$. In fact, $\phi = \frac{\pi}{4}$ ensures that clipping does not occur even for $A$ slightly in excess of $1/N$; the effect being profound for higher $N$. Beyond that $A$, the PAPR begins to increase until it reaches a maximum and then tapers off. This observation is against the expectation that the PAPR would decrease, since the onset of clipping restricts the amplitude of $\xB(t)$ to one while $\expect\{\xhat^2(t)\}$ keeps increasing, as seen before.

\item The above contradiction can be explained as follows. While DAC clipping limits the amplitude of $\xB(t)$ to one, the corresponding RF waveform peaks when $\xIbar(t) = \xQbar(t) = 1$ simultaneously, which occurs afterward\footnote{As evident in \eqref{xi}, the complexity of $\xB(t)$ prevents us from determining these time instants theoretically, and so we rely on simulations.} at higher $A$. Then onwards the peak power value of the RF waveform stays fixed at four while $\expect\{\xhat^2(t)\}$ keeps increasing and thereby reducing the PAPR. As $A$ goes on increasing further, $\expect\{\xhat^2(t)\}$ converges to two, and thus from \eqref{PAPR2}, the PAPR converges to two.

\begin{figure*}[t]
	\centering
	\subfloat[test-bed for measuring $\etaDCtoDC$]{\includegraphics[width=0.33\textwidth]{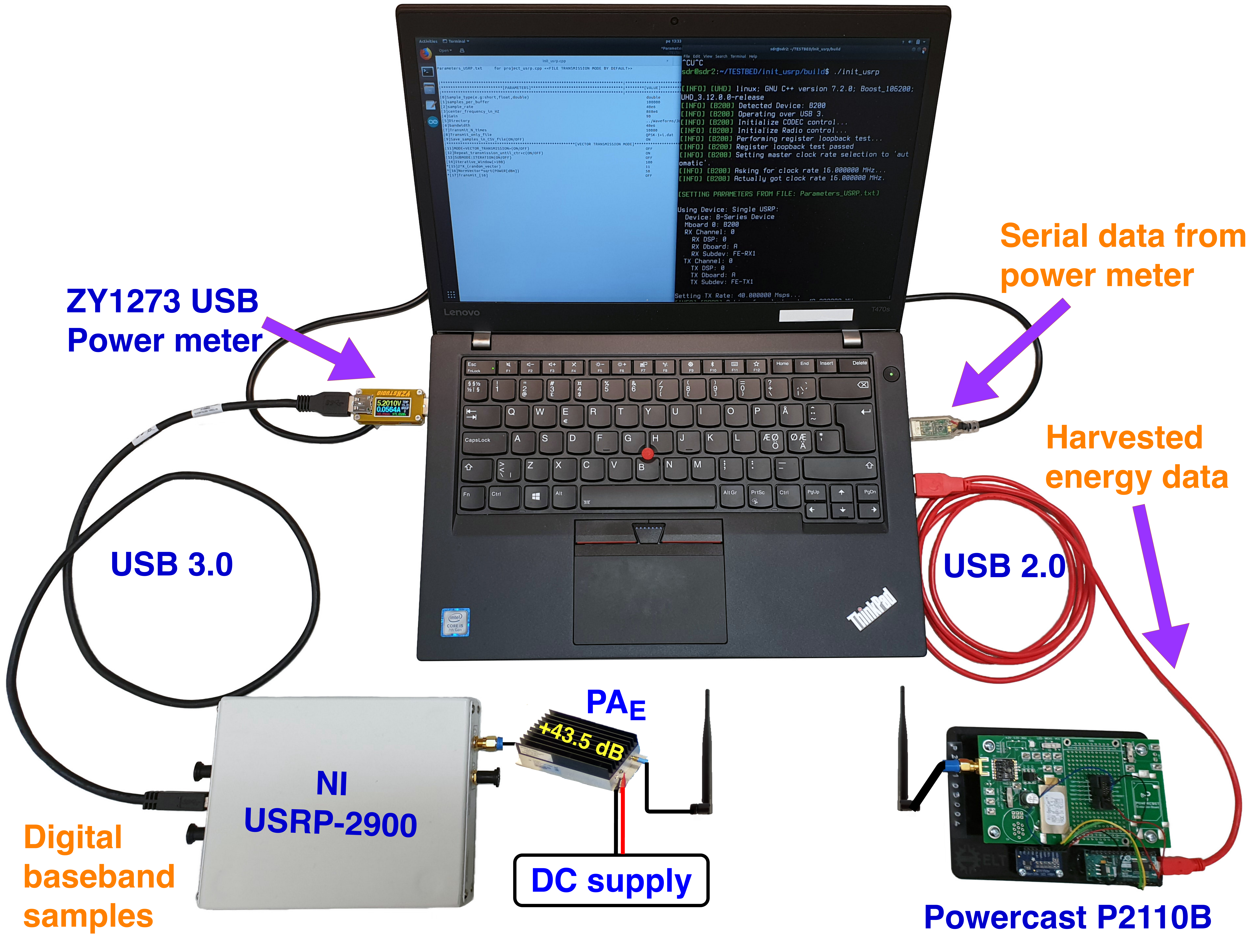}\label{fig:test-bed_a}} 
	\subfloat[test-bed for measuring $\PRFout$, $\etaDCtoRF$ and the EVM]{\includegraphics[,width=0.33\textwidth]{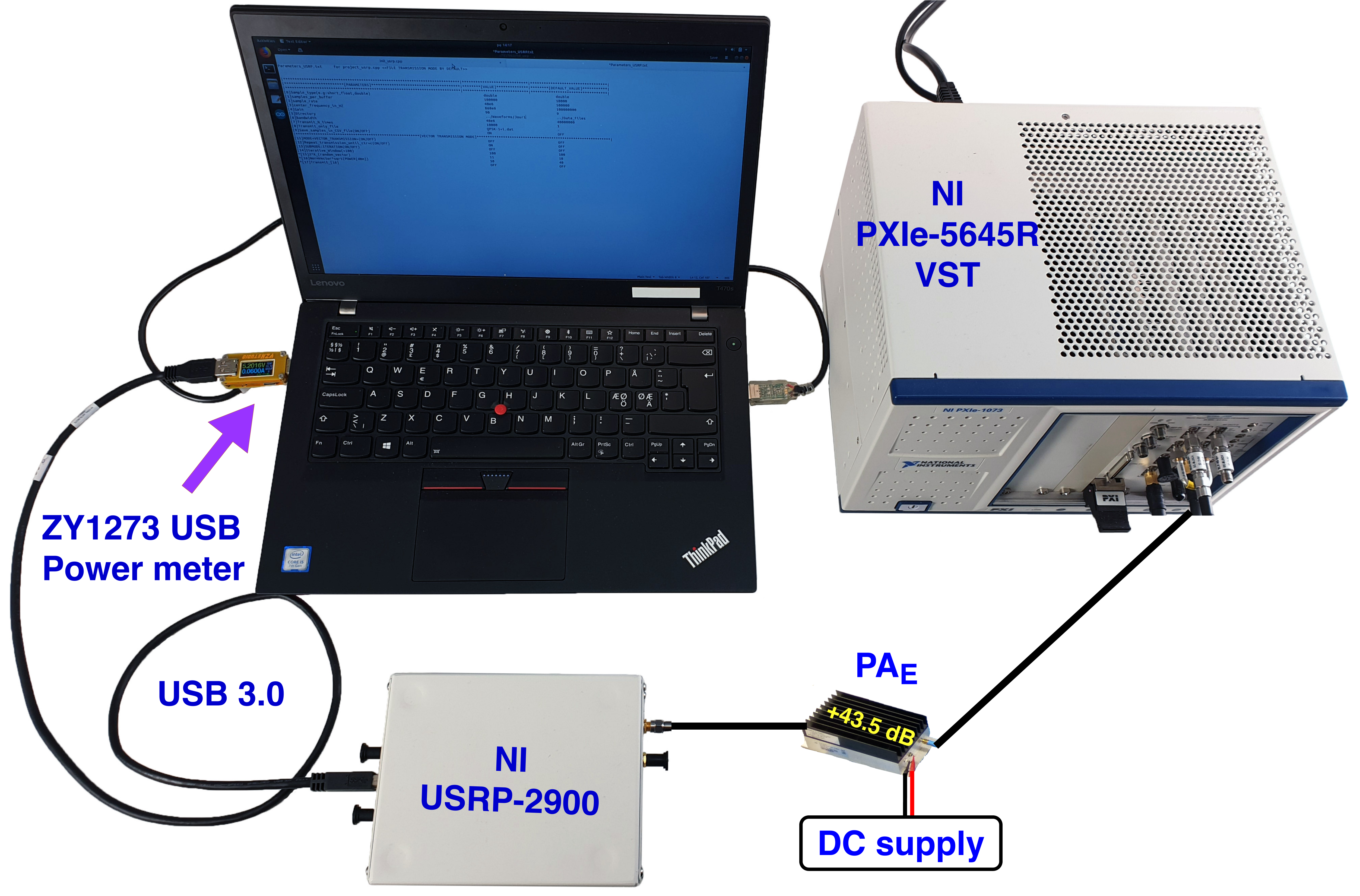}\label{fig:test-bed_b}}
    \subfloat[test-bed for measuring $\PRFin$]{\includegraphics[width=0.33\textwidth]{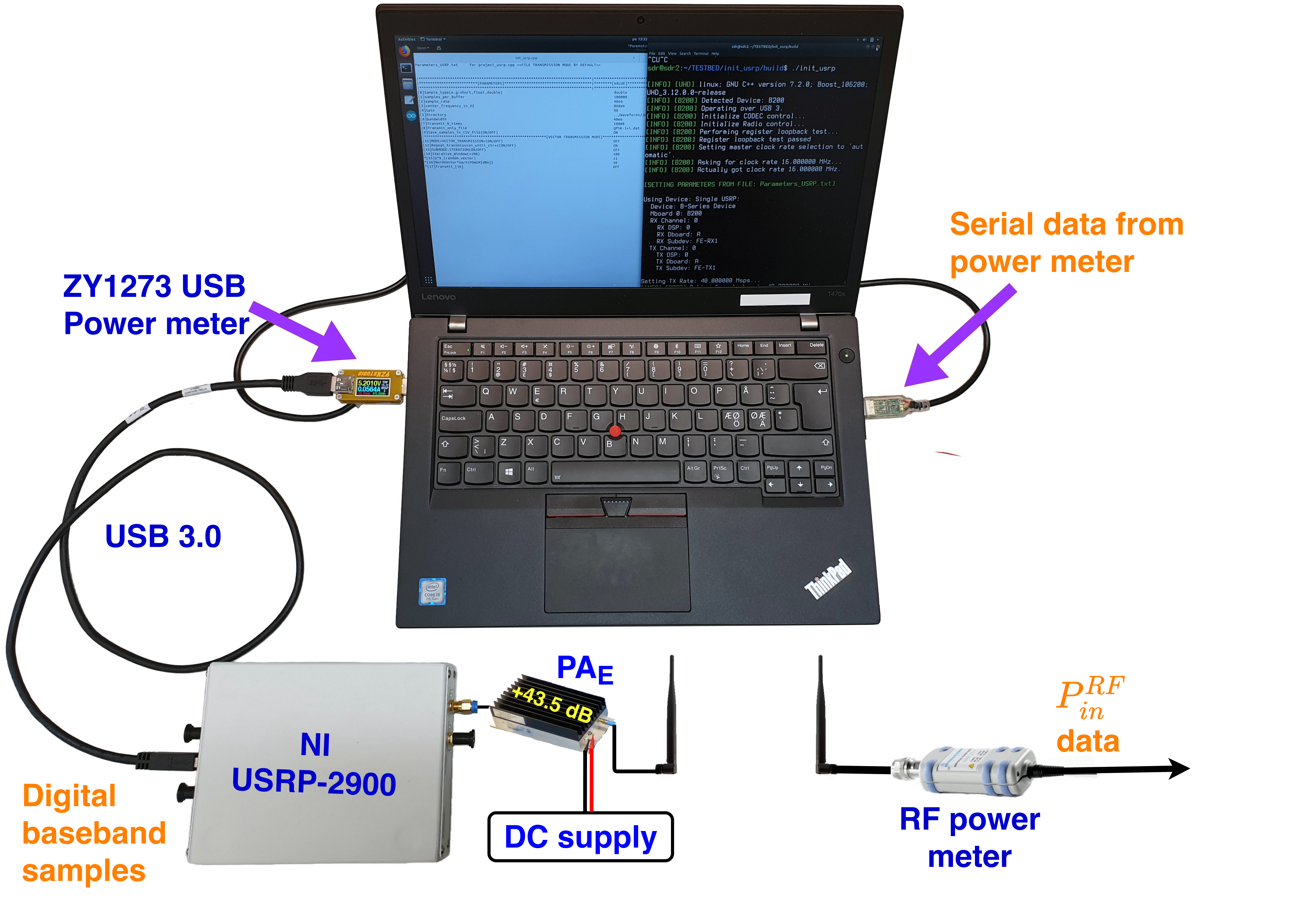}\label{fig:test-bed_c}}
	\caption{\label{fig:test-bed}The USRP-based test-beds for evaluating  the end-to-end, transmitter and channel efficiencies in RF WPT, and for assessing the quality of the transmitted RF signal in terms of the EVM.}
\end{figure*}

\item Next, we evaluate the impact of DAC clipping and quantization on the EVM of the $x(t)$. We adopt the standard definition of EVM, given as 
\begin{equation}\label{evm_multisine}
    {\rm EVM} = \sqrt{\frac{\expect\{|x(t)-\bar{x}(t)|^2\}}{\expect\{|x(t)|^2\}}}.
\end{equation}
The results are shown in Fig.~\ref{fig:DAC_efficiency_c}, where we have plotted the EVM in logarithmic scale for different $N$ and varying $A$. We observe that while $A \le \frac{1}{N}$, the EVM is very small. In this region, the only source of error is the quantization process. The low EVM suggests that  a typical 12-bit DAC is capable of handling the fast varying tails of a multisine signal. Now, for $A > \frac{1}{N}$, the sharp increase in the EVM shows how detrimental is the clipping caused by the DAC. Here again, comparison based on same input average power reveals that the EVM performance degrades with increasing $N$.
\end{enumerate}

Based on these results, we conclude that, before amplification by the $\pai$, a single-tone signal has both the highest average RF power and the lowest EVM. Moreover, adding more sinusoids to achieve a high-PAPR signal tends to lower the average RF power and also degrades the EVM performance due to the clipping effect of the DAC.  While lowering $A$ helps to avoid clipping, it eventually results in lower average transmit power and becomes detrimental to $\etaDCtoRF$ (and thus $\etaDCtoDC$). Most research works referenced in \cite{reference} consider that the average power of all the multisine waveforms is kept equal for a fair comparison. This does not help much in practice either since digital waveforms with higher PAPR but same average power would be clipped by the DAC. The resultant analog baseband waveforms would neither have the same average power nor the same PAPR.

The analog RF waveform generated by the USRP is amplified by the $\pai$ and an external PA ($\pae$) before transmission over a wireless medium. The detrimental impacts of PAs on high-PAPR signals are well known and hence avoided here for brevity. While operating the $\pai$ with a large back-off assures linear amplification, it also results in lower average radiated RF power. We shall see in Section~~\ref{sec:experiments} that this eventually leads to lower $\etaDCtoRF$, and thus lower $\etaDCtoDC$.

%%%%%%%%%%%%%%%%%%%%%%%%%%%%%%%%%%%%%%%%%%%%%%%%%%%%%%%%%%%%%%%%%%%%%%%%%%%%%%%%%%%%%%%%%%%%%%%%%%%%%%%%%%%%%%%%%%%%%%%%%%%%%%%%%%%%%%%%%%%%%%%%%%%%%%%%%%%%%%%%
\section{Test-bed for RF Wireless Power Transfer}\label{sec:test-bed}

We now explain the test-bed developed for this work. The primary purpose of our test-bed is to evaluate $\etaDCtoDC$ of RF WPT for digital baseband test waveforms. Moreover, we aim to analyze the impact of different stages of power transfer---RF signal generation, channel propagation, and energy harvesting---on $\etaDCtoDC$. Employing an SDR for transmission makes our research worthwhile mainly for SWIPT. Thus, we also  observe the impact of transmitter imperfections on the quality of the transmitted signal by evaluating EVM.

As the test-bed's SDR, we have opted for NI USRP-2900, where the $\pai$ has a tunable gain that can be set with the USRP gain setting ($G$) value, and the maximum output power is around $20$~dBm. We also incorporate a $\pae$, which ensures that the RF signal at the receiver input exceeds the detection threshold ($-15$~dBm) of the energy harvester. An overview of the test-bed is shown in Figs.~\ref{fig:sys_a} and~\ref{fig:test-bed}. While the current setup in Fig.~\ref{fig:test-bed} appears to be more deconstructed for experimental testing, we did so on purpose for evaluating power transfer efficiencies as well as the EVM for test waveforms.\footnote{ A true SWIPT system that characterizes both power and data is the next step in our research, wherein we intend on testing both the time-splitting (TS) and power-splitting (PS) approaches at the receiver.    In comparison with Fig.~\ref{fig:sys_a}, a SWIPT system would incorporate an additional block at the receiver-side which comprises an RF switch (resp. TS) or a power splitter (resp. PS). The VST would then be used for information decoding and aiding in computing the bit error rate of the communication link. Alternatively, a SWIPT system can comprise a separate communication receiver and a few energy harvesting sensors. While employing a single digital baseband waveform, we can examine the efficacy of data communication with the communication receiver while monitoring the end-to-end WPT efficiency of the sensors.} To the best of the authors' knowledge, our test-bed is the first Linux-based GNU Radio RF WPT setup in a C++ development environment. The hardware configurations of the equipment used in our test-bed are summarized in Table~\ref{table:1}.

\subsection{Setup for Evaluating the End-to-End Efficiency $\etaDCtoDC$}
While measuring $\etaDCtoDC$, the transmitter section consists of a computer, a USB power meter and an SDR. The computer feeds the digital baseband samples to the SDR through a USB 3.0 interface. Moreover, the computer supplies DC power ($\PDCUSRP$) to the SDR, which is measured by the USB power meter placed in series between the computer and the SDR. The USB power meter communicates the real-time power consumption data over a USB 2.0 interface to the computer, where it is logged for further analysis. The SDR generates the modulated RF signal, which is further amplified by the $\pae$ to $\PRFout$, for wireless transmission through a whip antenna. A separate bench DC supply powers the $\pae$ and reports the power consumption data ($\PDCPA$) to the computer. So, the total power consumption at the transmitter side is $\PDCin=\PDCUSRP+\PDCPA$.

The receiver side comprises a multi-band whip antenna, an off-the-shelf\footnote{We use off-the-shelf components in our experiments to attain a fair idea about the viability of RF WPT in real-world applications.} RF energy harvester board, with its supporting circuitry, and a computer. The harvester is equipped with a diode-based rectifier that converts the incident RF energy ($\PRFin$) into DC. It is accompanied by a discrete analog-to-digital converter (ADC) and a microcontroller that continuously measures the instantaneous DC voltage across an onboard resistor $R$. The microcontroller continuously reports the instantaneous DC power data to the computer over a USB 2.0 interface. The computer logs this information to calculate the average harvested DC power ($\PDCout$) for each transmission interval. Based on these measurements, we compute $\etaDCtoDC = \PDCout / \PDCin$. The setup for measuring $\etaDCtoDC$ is shown in Fig.~\ref{fig:test-bed_a}.   

\subsection{Setup for Evaluating the Transmitter Efficiency $\etaDCtoRF$ and the EVM of the Transmitted Signal}\label{sec:test-bed-tx_eff}
While measuring $\etaDCtoRF$, the transmitter section up to the $\pae$ remains unchanged w.r.t.\ the above one, and we obtain information on $\PDCin$ as explained before.  The receiver side now comprises a VST and a computer. The $\pae$ output is connected to the VST input by a coaxial cable. The VST captures the RF waveform and provides the digital IQ samples to the computer for evaluating the EVM and $\PRFout$. Based on these measurements, we compute $\etaDCtoRF = {\PRFout}/{\PDCin}$. The setup for measuring $\etaDCtoRF$ is shown in Fig.~\ref{fig:test-bed_b}. 

\subsection{Setup for Evaluating the Channel Efficiency $\etaRFtoRF$}
In order to compute $\etaRFtoRF$, the measurement data for $\PRFout$ is obtained as explained above. To obtain the measurement data for $\PRFin$, the receiver side comprises a multi-band whip antenna, an RF power meter and a computer. The  RF power meter measures $\PRFin$ at the antenna port, and reports the data to the computer. Thus, we compute $\etaRFtoRF=\PRFin/\PRFout$. The setup for measuring $\etaRFtoRF$ is presented in Figs.~\ref{fig:test-bed_b} and \ref{fig:test-bed_c}.

\subsection{Differences With Similar Test-beds}
Our RF WPT test-bed implementation differs from \cite{bruno_arxiv,Clerckx_rectenna} in being just a simple open-loop system. The practical implementations in \cite{bruno_arxiv,Clerckx_rectenna} comprise CSI feedback to the transmitter to optimize the weights of the sinusoids in a multisine waveform, which results in enhanced energy harvesting performance. However, the CSI feedback in the aforementioned reference implementations is retrieved over a wired channel and not wirelessly. In addition, the channel estimation is performed at the same VST as the transmitter. We believe that, although they represent seminal work as good research exercises, the manner in which the feedback was obtained may not be very practical in certain cases.

The targets of WPT/SWIPT are typically simple sensor nodes that work on few hundred microwatts' power or even much less. In a practical scenario, such nodes may not be able to perform complex tasks such as channel estimation and transmit the CSI back to the transmitter in real time with their limited energy reserves and computing capability. In view of such practical constraints, it is reasonable to pursue the approach of operating over a narrow bandwidth (flat-fading channel) with uniform power allocation for the sinusoids of the multisine signal, which then eliminates the need for CSI feedback and also performs equally well as the adaptive designs  \cite{bruno_arxiv,BC1}. However, this simplicity in operation comes at the cost of a significant performance loss when having multi-antenna transmitters or a frequency-selective channel~\cite{bruno_arxiv}.

\begin{table}[t]
\renewcommand{\arraystretch}{1.2}
\caption{Hardware configuration for the test-bed.}
\begin{center}
\begin{tabular}{ | m{2.5cm} | m{5.2cm}| } 
\hline
\textbf{Component} & \textbf{Product details} \\ 
\hline
Computer & Lenovo ThinkPad T470p, 32GB RAM, Core i7 processor, Linux  OS, GNU C++ compiler\\ 
\hline
SDR & National Instruments USRP-2900 \\ 
\hline
USB power meter & YZXStudio ZY1273 \\ 
\hline
External PA & Mini-Circuits ZHL-4240+ \\
\hline
Antenna & Siretta Delta 6A multi-band whip antenna\\
\hline
Network analyzer & Anritsu S820E \\ 
\hline
RF energy harvester & Powercast P2110B \cite{datasheet_P2110B} \\ 
\hline
ADC & Texas Instruments 16-Bit ADS1115 \\ 
\hline
Microcontroller & Arduino Nano (Microchip ATmega328) \\ 
\hline
VST & National Instruments PXIe-5645R \\
\hline
RF power meter & Rohde \& Schwarz NRP-Z11\\
\hline
\end{tabular}
\label{table:1}
\end{center}
\vspace{-12pt}
\end{table}

\begin{table}[t]
\renewcommand{\arraystretch}{1.2}
\caption{Operational parameters for the experiments.}
\begin{center}
\begin{tabular}{ | m{4cm} | m{3.7cm}| } 
\hline
\textbf{Parameter} &  \textbf{Details} \\ 
\hline
Frequency band & $863-873$~MHz (unlicensed) \\ 
\hline
Carrier frequency & $\fc=868$~MHz \\ 
\hline
Sampling rate & $40$~MHz \\ 
\hline
Baseband fundamental frequency & $f_0 = 200$~kHz\\ 
\hline
Number of tones in multisine & $N\in \{1,2,4,8,16\}$ \\ 
\hline
PAPR & $\begin{cases}
        2,   & N=1 \\
        [4, 2N] ~(0.5 \textrm{ steps}),   & N > 1
       \end{cases}$ \\
\hline
USRP gain setting & $G \in [40, 57]$~dB\\
\hline
$\pae$ gain  & $43.5$~dB\\
\hline
Resistive load& $R=286~\Omega$\\
\hline
\end{tabular}
\label{table:2}
\end{center}
\vspace{-12pt}
\end{table}

%%%%%%%%%%%%%%%%%%%%%%%%%%%%%%%%%%%%%%%%%%%%%%%%%%%%%%%%%%%%%%%%%%%%%%%%%%%%%%%%%%%%%%%%%%%%%%%%%%%%%%%%%%%%%%%%%%%%%%%%%%%%%%%%%%%%%%%%%%%%%%%%%%%%%%%%%%%%%%%%
\section{Experimental Data-Driven Analysis of\\RF WPT Efficiency}\label{sec:experiments}

\begin{figure}[t]
\begin{center}
    \includegraphics[width=\linewidth]{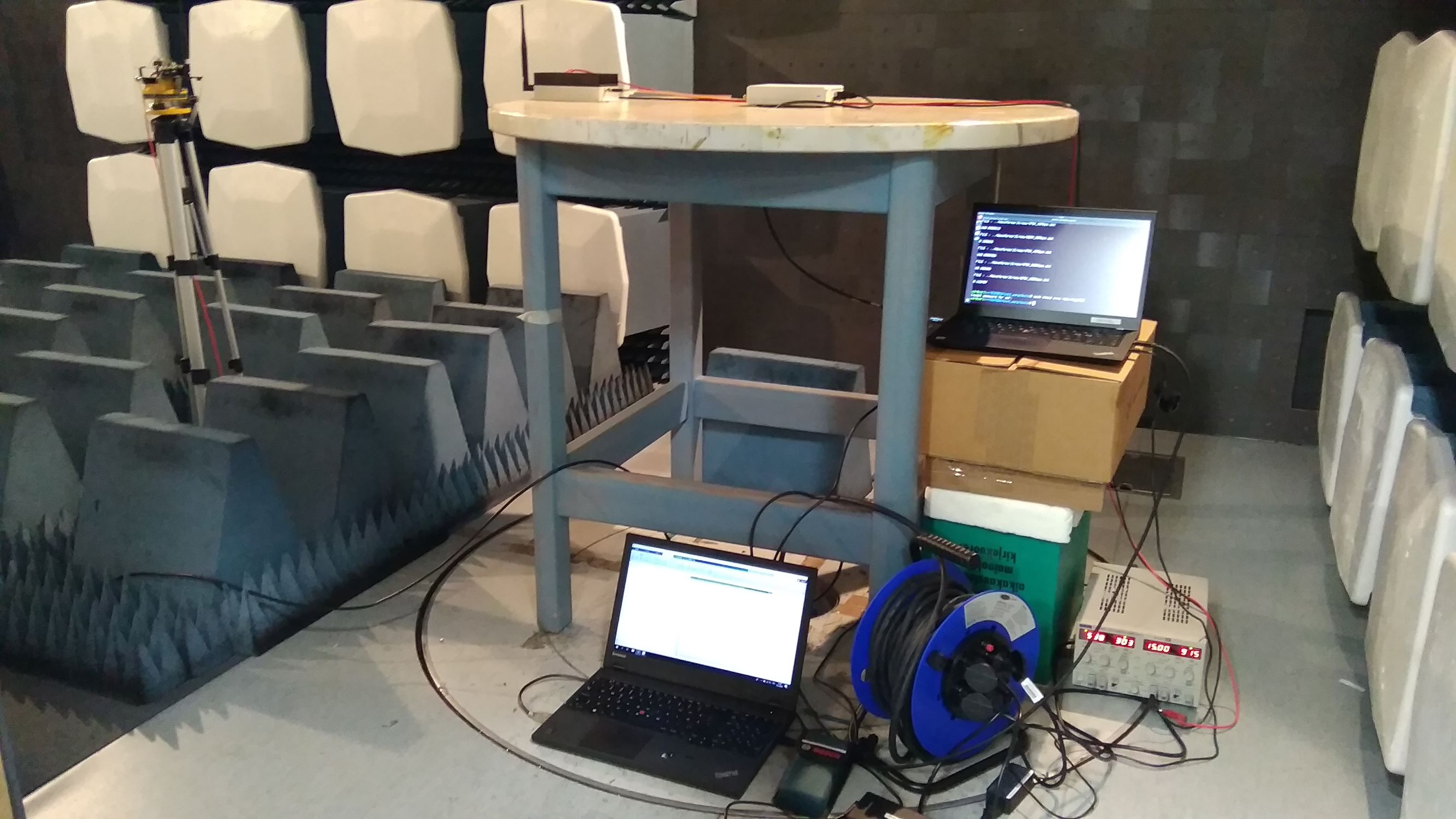}
    \caption{Anechoic chamber setup for RF WPT experiments.}
    \label{fig:chamber_setup}
\end{center}
\end{figure}
\begin{figure}[t]
    \includegraphics[width=\linewidth]{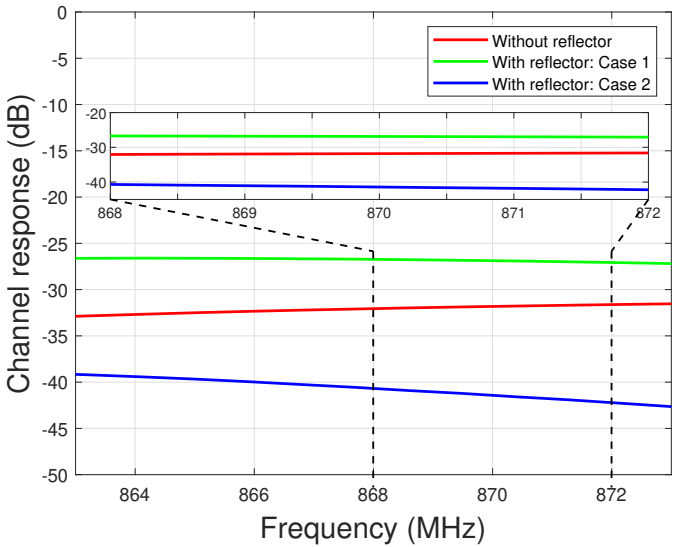}
    \caption{Channel response in the frequency band of operation.}
    \label{fig:channel_response}
\end{figure}

In this section, we present the various experimental results and corroborate the observations with theory. The parameters for the experiments are summarized in Table~\ref{table:2}. The range of $G$ is determined by the setup: upper limit by the input threshold power for the $\pae$ ($-5$~dBm), and lower limit by the minimum acceptable $\PDCout$.  Our harvester board comprises an onboard resistor $R=286$~$\Omega$. Thus, it is important to note that all the measurement results presented here are valid for resistance values of the order of a few hundred ohms. The harvester board also comprises a DC-to-DC boost converter, which we do not employ in this work as we focus on accessing  the rectifier performance.\footnote{The RF-to-DC efficiency of multisine waveforms for varying loads in the presence of a DC-to-DC boost converter has been studied in \cite{Bruno_DC-DC_Optimal_Operation_of_Multitone_Waveforms_in_Low_RF-Power_Receivers}.}

\begin{figure}[t]
    \subfloat[$D=1$~m, Anechoic (without reflector) ]{\includegraphics[width=0.31\linewidth]{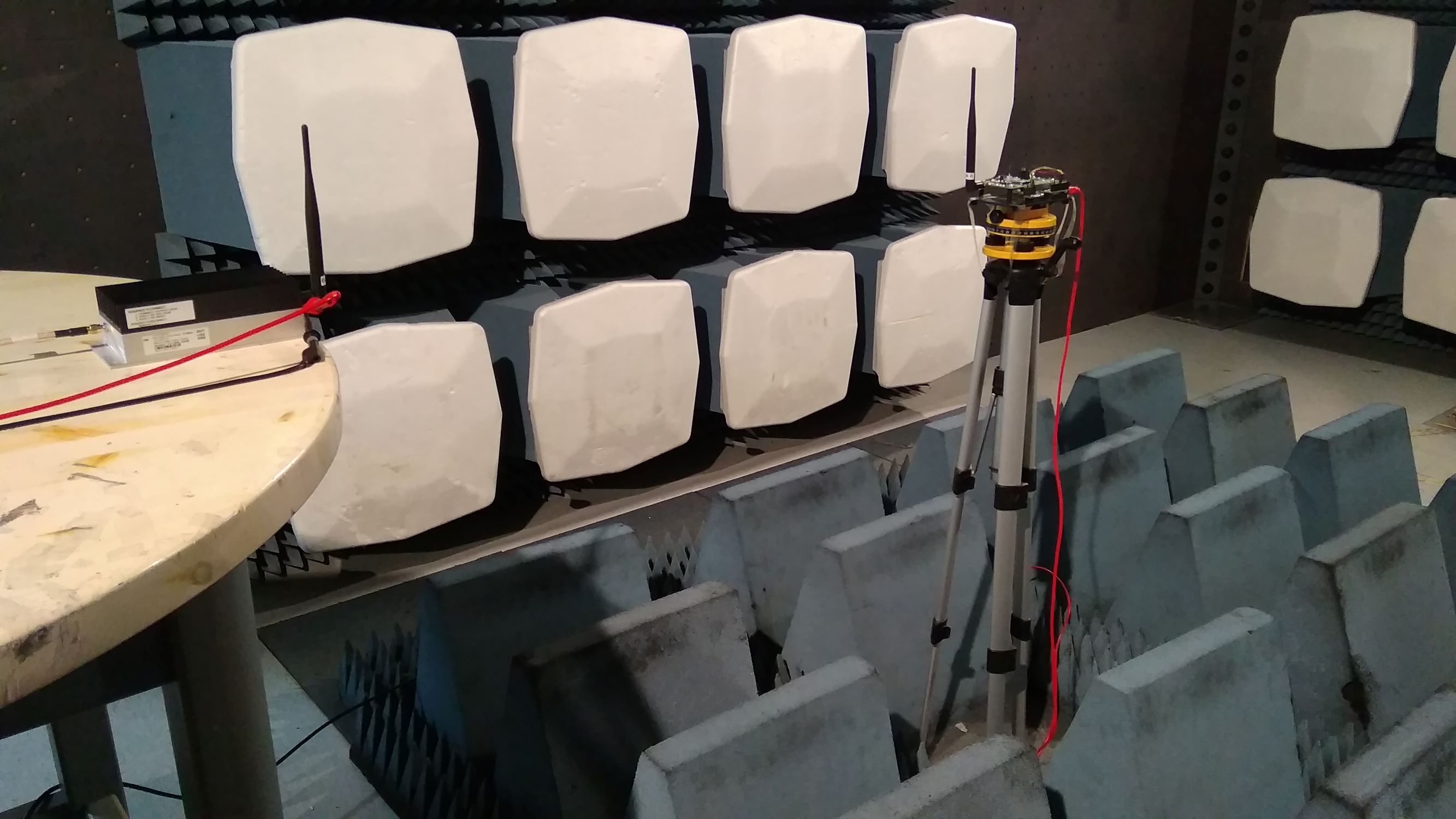}\label{fig:chamber_1m_no_reflector}}
    \hspace{5pt}
    \subfloat[$D=1$~m, Case 1 (with reflector)]{\includegraphics[width=0.31\linewidth]{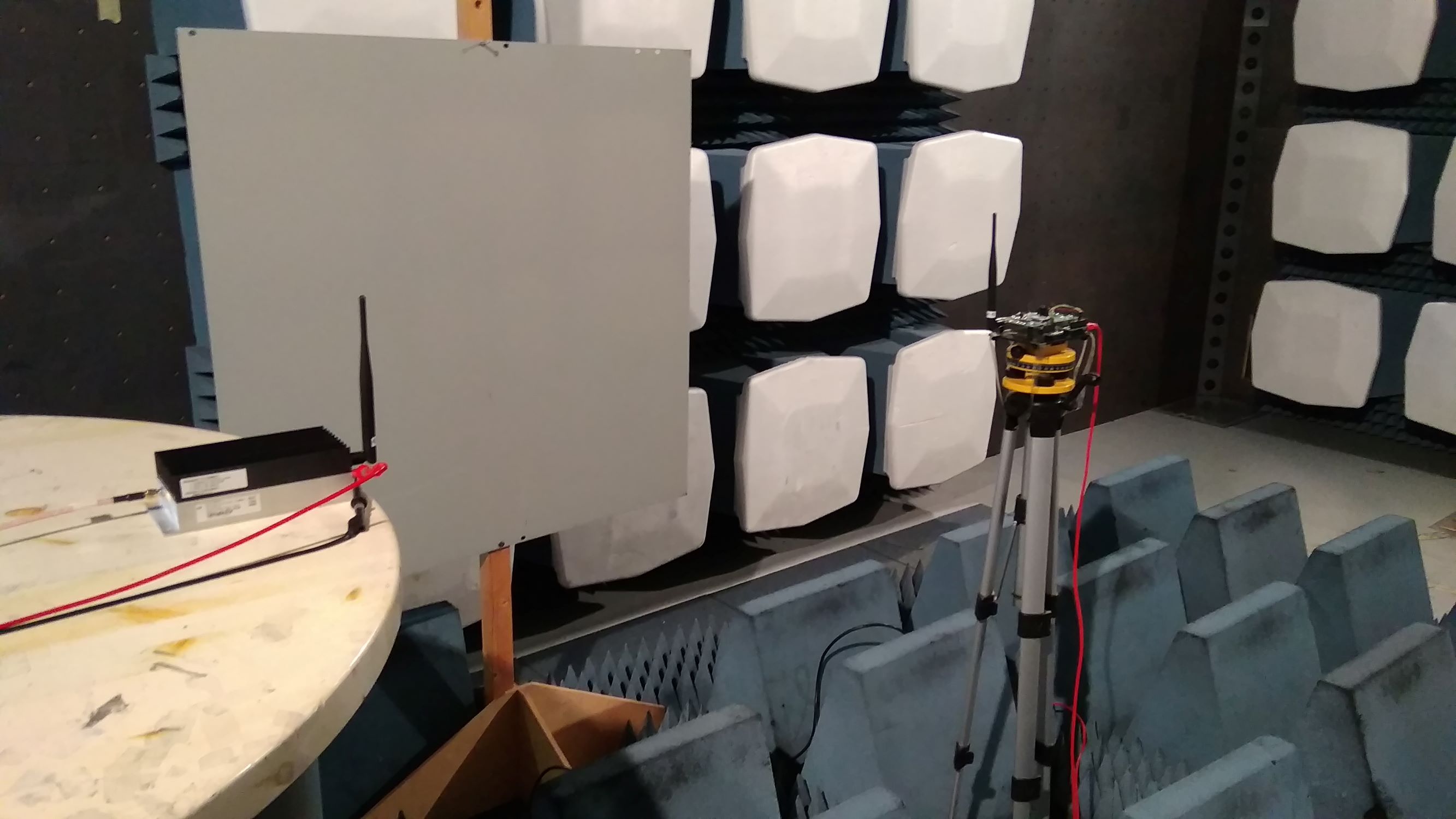}\label{fig:fig:chamber_1m_near}}
    \hspace{5pt}
    \subfloat[$D=1$~m, Case 2 (with reflector)]{\includegraphics[width=0.31\linewidth]{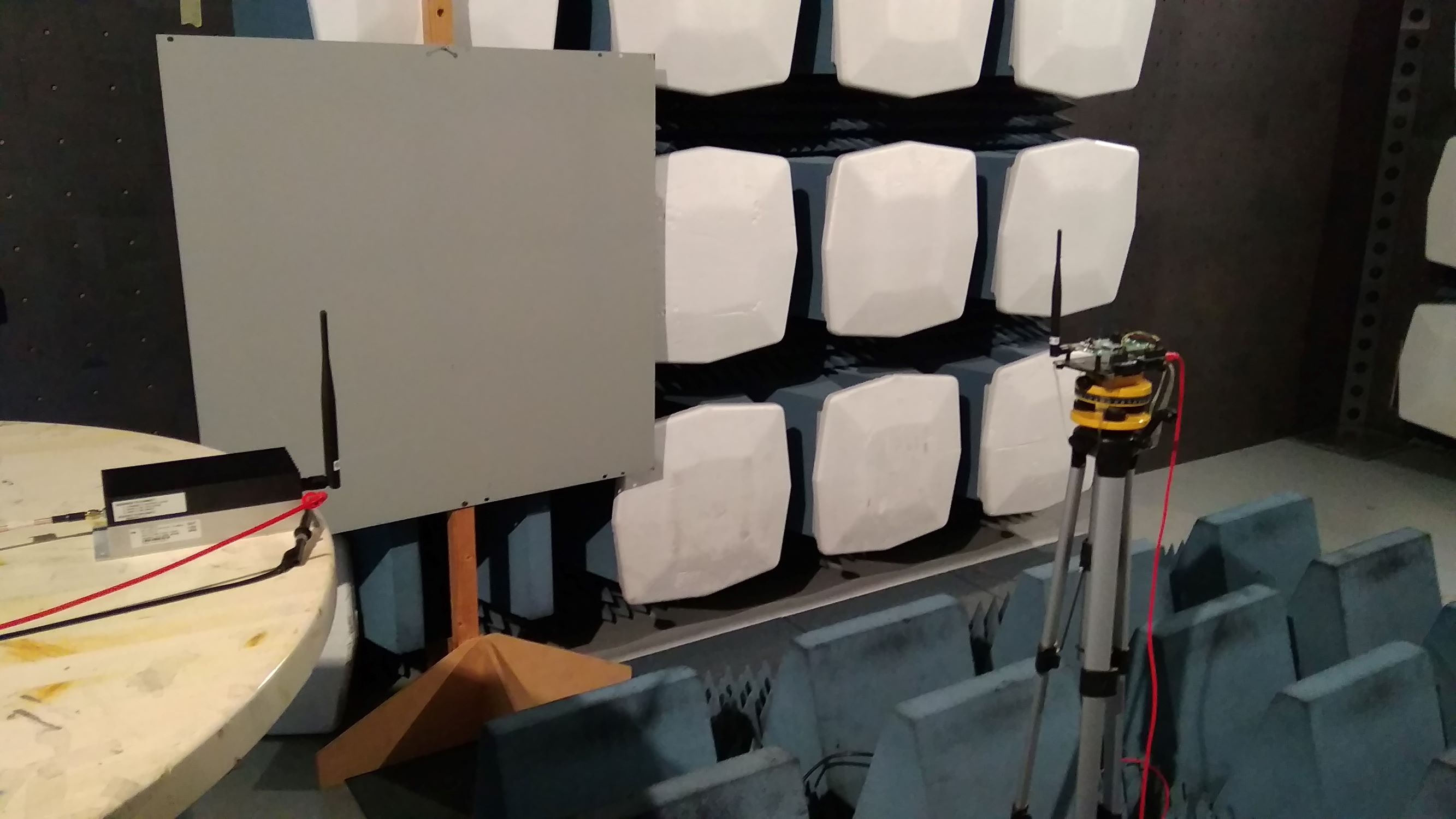}\label{fig:chamber_1m_far}}\\
    \subfloat[$D=2$~m, Anechoic (without reflector)]{\includegraphics[width=0.31\linewidth]{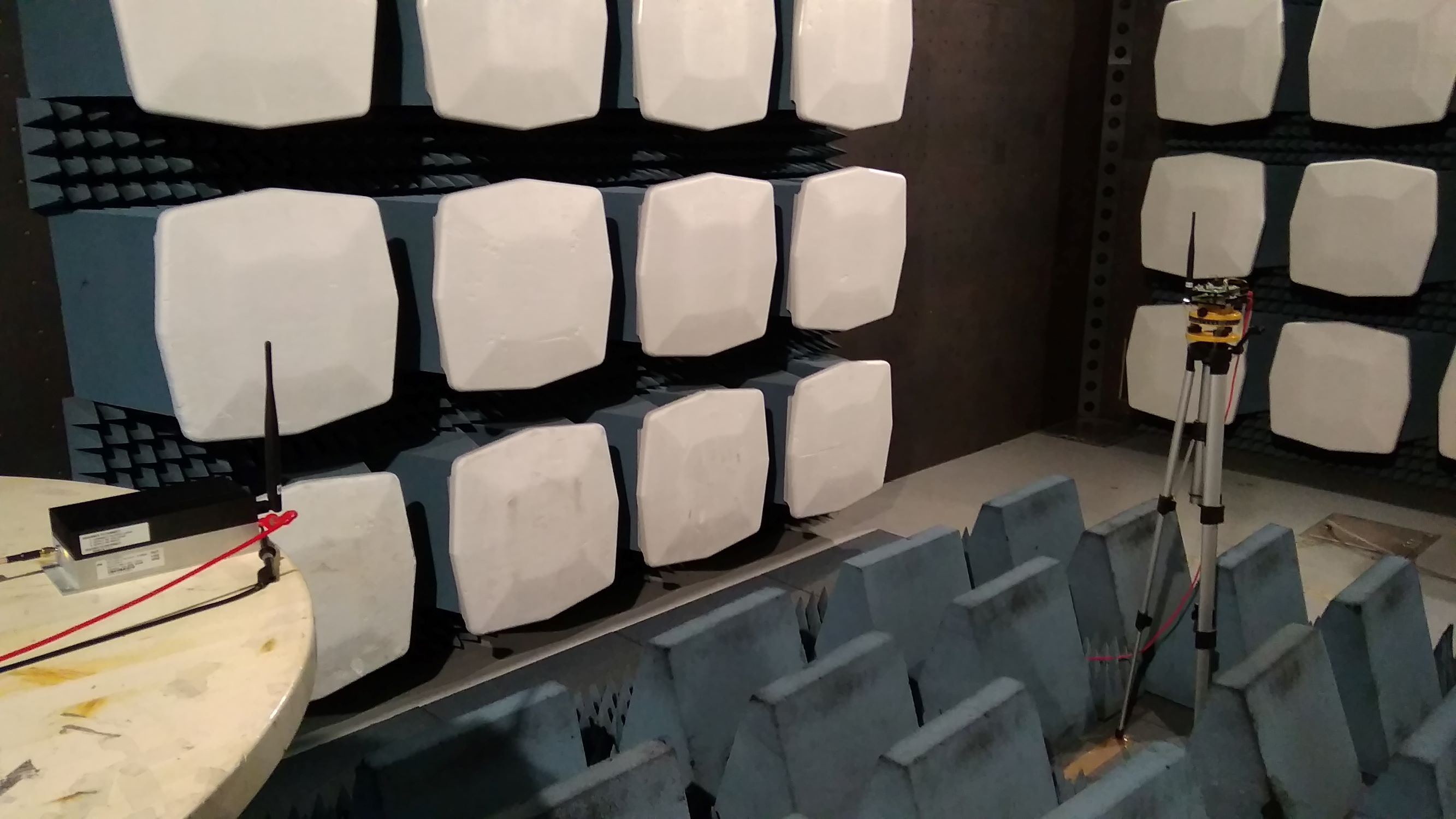}\label{fig:chamber_2m_no_reflector}}
    \hspace{5pt}
    \subfloat[$D=2$~m, Case 1 (with reflector)]{\includegraphics[width=0.31\linewidth]{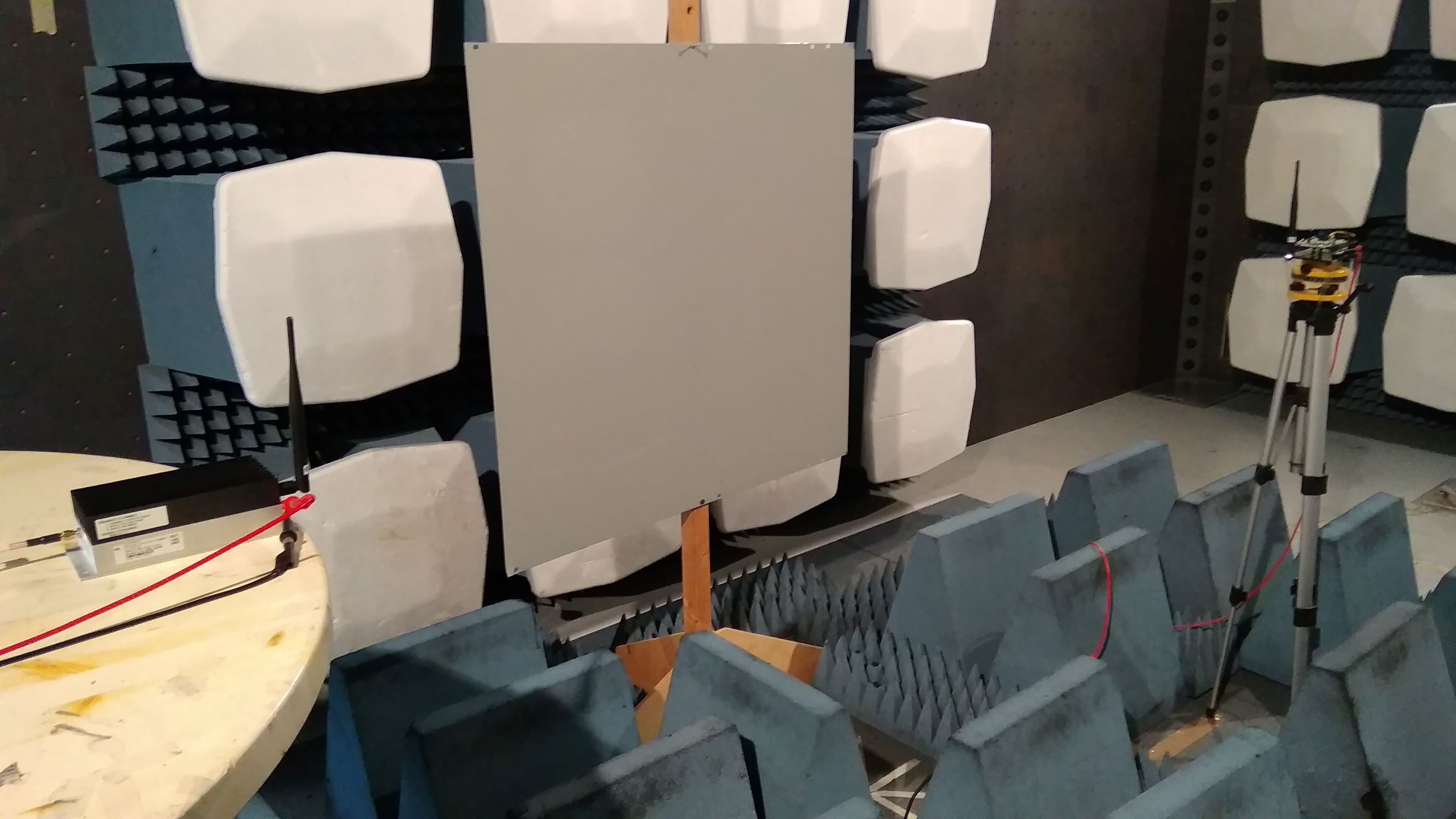}\label{fig:fig:chamber_2m_near}}
    \hspace{5pt}
    \subfloat[$D=2$~m, Case 2 (with reflector)]{\includegraphics[width=0.31\linewidth]{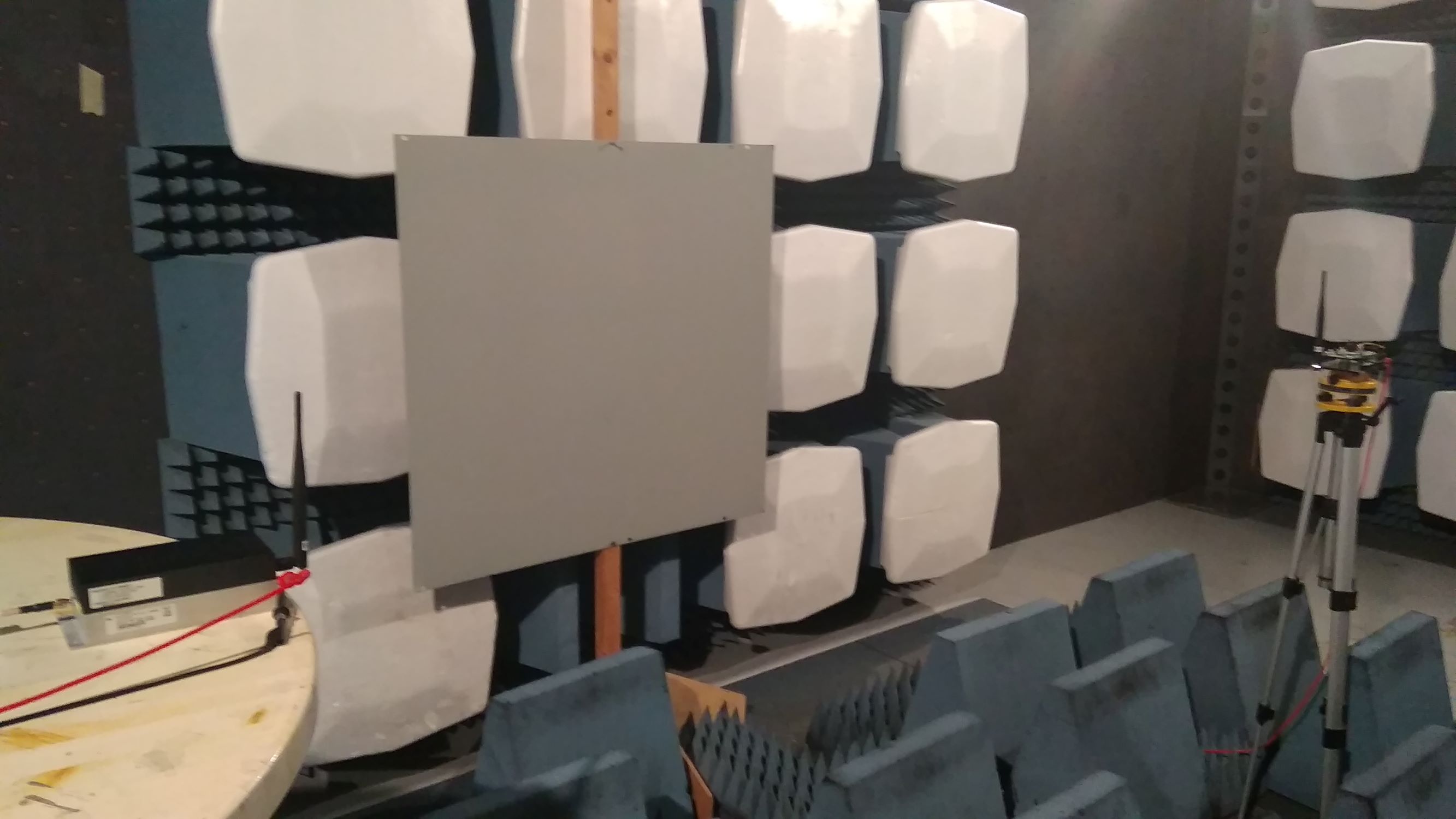}\label{fig:chamber_2m_far}}
    \caption{Anechoic chamber setup for measuring $\etaDCtoDC$. The position of the reflecting surface and the transmitter--receiver distance $D$ are varied to experiment with varying channel conditions.}
    \label{fig:chamber_etaDCtoDC}
\vspace{-2mm}\end{figure}

All the wireless experiments were performed in an anechoic chamber to avoid unnecessary interference with the neighboring licensed GSM band. We also placed a reflector within the anechoic chamber to simulate a real wireless multipath propagation environment with two cases. Moreover, the transmitter--receiver distance ($D$) and the location of the reflector were also varied to observe different levels of attenuation. The experiment setup is shown in Fig.~\ref{fig:chamber_setup}.

\begin{figure*}[t]
	\subfloat[PAPR reduced due to clipping when $A=1$]{\includegraphics[width=0.32\textwidth]{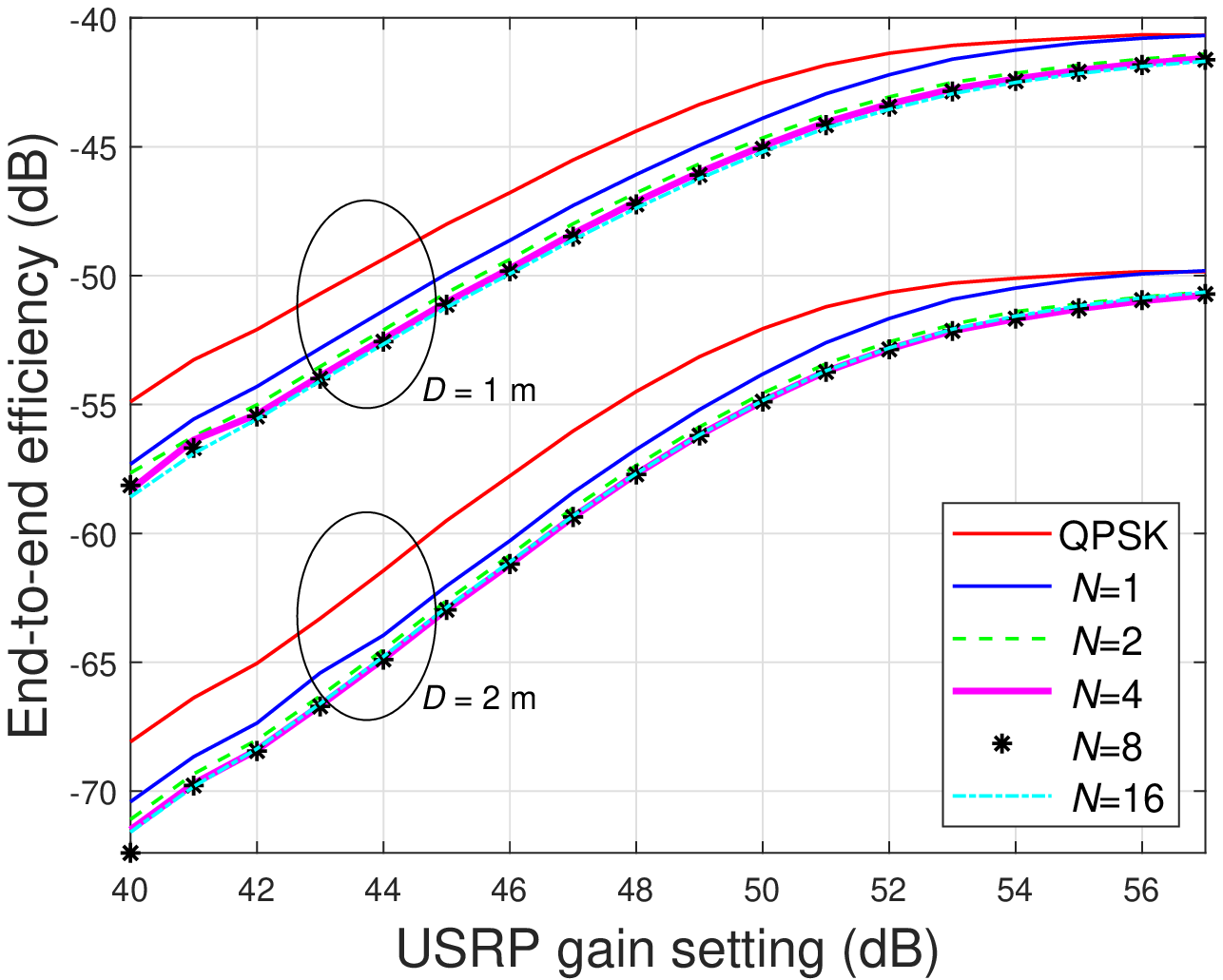}\label{fig:e2e_eff_a}} \hspace{5pt}
	\subfloat[PAPR reduced due to clipping when $A=1/\sqrt{N}$ and $A=1/N$]{\includegraphics[,width=0.32\textwidth]{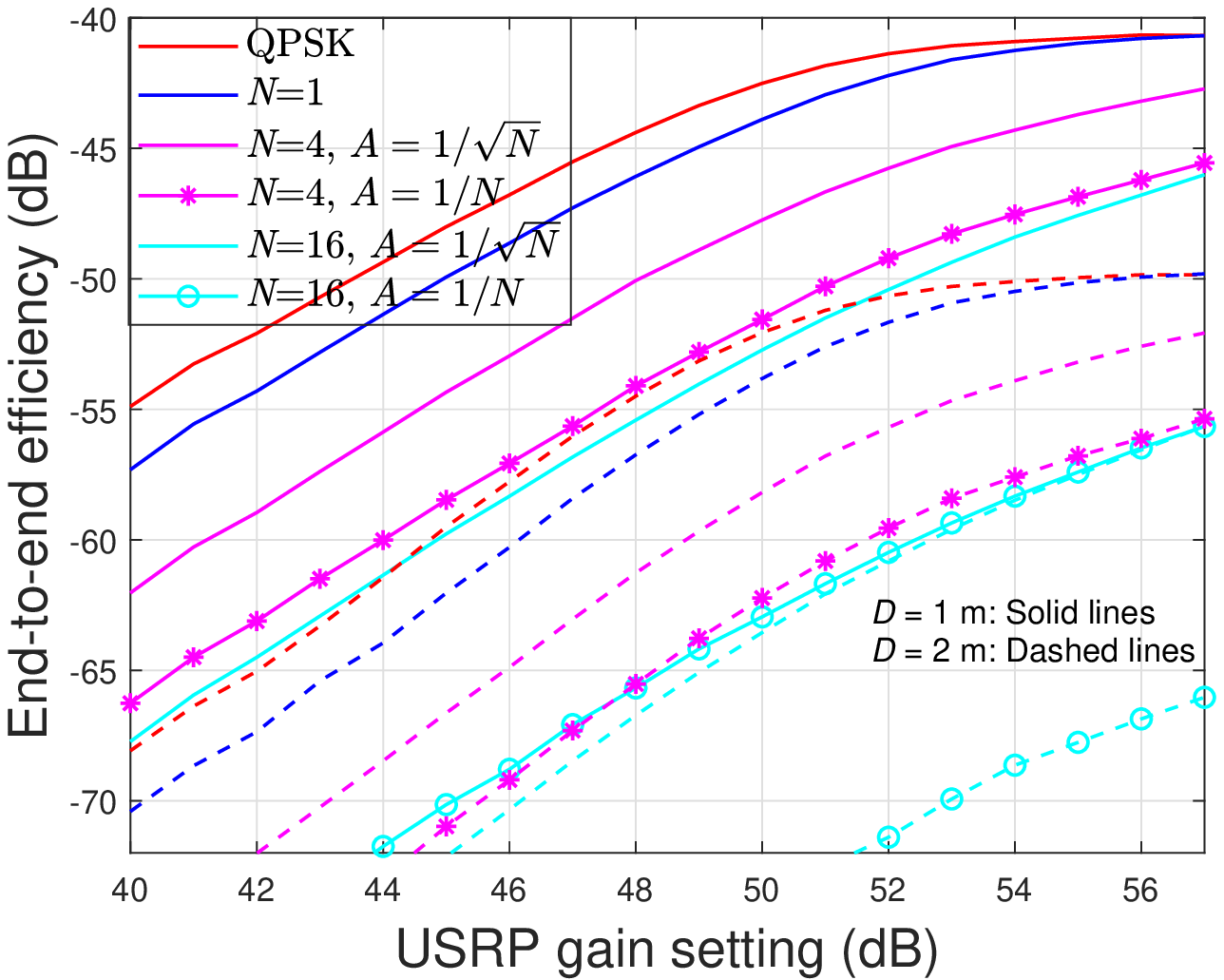}\label{fig:e2e_eff_b}} \hspace{5pt}
    \subfloat[varying PAPR with $N=4$]{\includegraphics[width=0.32\textwidth]{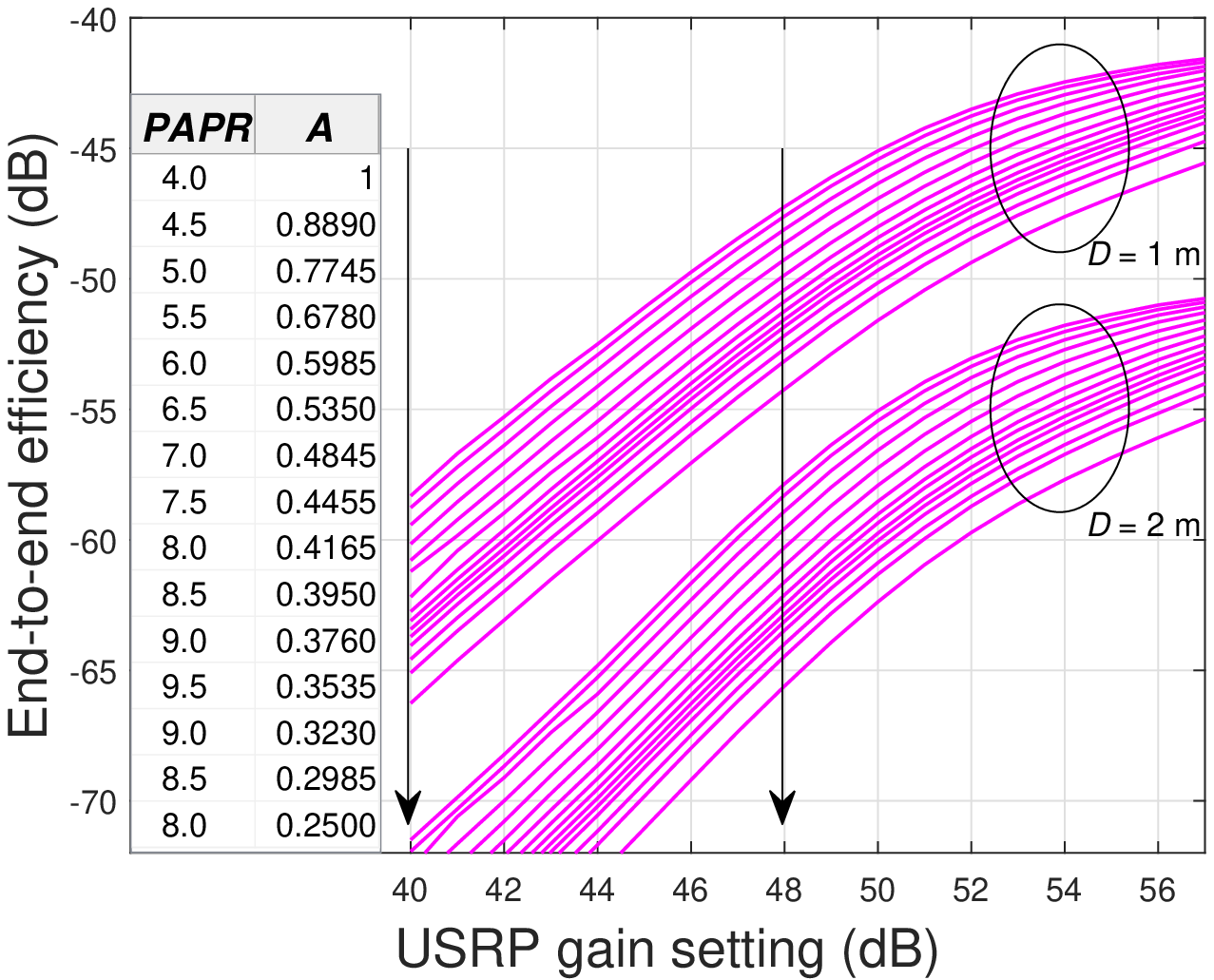}\label{fig:e2e_eff_c}}
	\caption{\label{fig:e2e_eff}Measurement results for the end-to-end WPT efficiency of  co-phased $N$-tone multisine waveforms by varying the amplitude $A$ and the USRP gain setting $G$. The QPSK reference waveform $x(t) = \pm 0.707 \pm 0.707j$ has an absolute magnitude of one, and thus is immune to DAC clipping.}
\end{figure*}
%%%%%%%%%%%%%%%%%%%

\subsection{Channel Sounding}\label{subsection:channel_sounding}
In order to determine the weighing coefficients of various components of a multisine signal, we need to determine the channel response in the frequency band of operation. For this purpose, we conducted channel sounding with the aid of a microwave network analyzer. The setup is similar to the one shown in Fig.~\ref{fig:chamber_setup} with the inclusion of the network analyzer. With $D=1$~m, we measured the channel response both in the absence and presence of a reflecting surface. In the latter case, the position of reflector was also varied, with respect to the transmitter and receiver, to ascertain the sensitivity of the channel response to it. The results of the channel sounding experiment are presented in Fig.~\ref{fig:channel_response}.

In Fig.~\ref{fig:channel_response}, the center curve represents the reference case with no reflecting surfaces around the transmitter and the receiver. The highest curve portrays the channel response when the reflector is deliberately placed at the best position with constructive scattering found between the transmitter and the receiver (Case $1$), while the lowest curve shows the channel response when the reflector is placed farther than the previous position (Case $2$) to find worst-case destructive scattering. Yet the primary observation is that the channel response varies only slightly in our frequency band of operation for all the three cases. Even with best effort, it was not possible to set up any significant frequency selectivity for the channel.

In accordance with other research works on this topic \cite{bruno_arxiv,bruno_Communications_and_Signals_Design_for_WPT}, we have also chosen similar values for $N$ in our experiments. Consequently, the maximum bandwidth, of undistorted multisine signals, in our experiments would be $3.2$~MHz ($868.2-871.4$~MHz). Accordingly, we have highlighted the channel response from $868-872$~MHz separately in Fig.~\ref{fig:channel_response}, where it can be seen to be flat-fading. Consequently, we can set the amplitude $A_n$ and the phase $\phi_n$ to be the same for all the sinusoids of a multisine signal, which concurs with the recommendations in \cite{bruno_arxiv,bruno_Communications_and_Signals_Design_for_WPT,Bruno_contrast_Robust_Wireless_Power_Receiver_for_Multi-Tone_Waveforms} for a flat-fading channel.

\subsection{End-to-End Efficiency of Multisine Waveforms versus $G$}\label{sec:e2e_eff}

We measured $\etaDCtoDC$ specifically for various multisine signals; their baseband expression is obtained by setting $A_{n} = A, \,f_{n} = n\,f_0$ and $\phi_{n} = \frac{\pi}{4}$ in \eqref{multisine}. Herein we present our observations for $N=1,2,4,8,$ and $16$. Each transmission lasted for around $10$~s and we computed the average $\etaDCtoDC$ of each waveform from three such transmissions. 

The measurement results are shown in Fig.~\ref{fig:e2e_eff}. As a reference case, 
we have chosen a quadrature phase-shift keying (QPSK) baseband signal (two bits per symbol) given by $x(t) = \xbar(t) = (\pm 1 \pm j)/\sqrt{2}$ that renders a bit rate of $1$~Mbps. We measured $\etaDCtoDC$ for six cases in total ($D=1$~m and $D=2$~m for the three cases shown in Fig.~\ref{fig:channel_response}); however for brevity, we only present herein the results for the two cases without the reflector. Moreover, it is clear from the plots in Fig.~\ref{fig:e2e_eff}  that the observations for the two propagation distances are similar, and differ only by a constant attenuation factor due to the channel being flat-fading. Thus, the forthcoming discussions are common for both the cases.

Fig.~\ref{fig:e2e_eff_a} depicts the end-to-end performance of the multisine signals when $A=1$. We observe that for $N \geq 2$, all the multisine signals yield almost similar efficiency, which is lower than the efficiency of a single-tone signal. The reference QPSK signal achieves the highest efficiency at all $G$ values. Thus with $A = 1$, adding more sinusoids gives no improvement in performance. This can be attributed to the heavy clipping at the DAC as $N$ increases, which diminishes the average RF power (before the $\pai$ stage) as observed in the simulation results of Fig.~\ref{fig:DAC_efficiency_a}.

Next, for a fair comparison, we measured $\etaDCtoDC$ when the baseband test waveforms have same average input power ($A = \frac{1}{\sqrt{N}}$) before the $\pai$ stage, and when they are unclipped ($A=\frac{1}{N}$). We observe in Fig.~\ref{fig:e2e_eff_b} that in both the cases as $N$ increases, $\etaDCtoDC$ goes on reducing as the high-PAPR multisine signals fare far worse than a single-tone signal. In the latter case, we observe that scaling down the amplitude to preserve the shape of the waveform only worsens the $\etaDCtoDC$ performance, since it reduces the average radiated RF power

Finally, for a given $N$, we examine the effect of varying the PAPR of the RF waveform (before the $\pai$ stage) on $\etaDCtoDC$. Here we present the observations for $N = 4$, wherein we vary the PAPR from $4$ to $8$ by successively decrementing $A$ from $1$ to $\frac{1}{4}$ and the results are shown in Fig.~\ref{fig:e2e_eff_c}. We observe that, as we try to increase the PAPR from four to $2N$, $\etaDCtoDC$ decreases and this is true at the whole range of $G$. This implies that even with the PAPR retained at $2N$, the unclipped multisine signal is not able to achieve higher $\etaDCtoDC$ even at higher $G$ values. On the contrary, it has the lowest efficiency at all $G$ values. From these observations, we can conclude that reducing the amplitude of the multisine signal to avoid clipping and then relying on the PAs for amplification may not be suitable while considering $\etaDCtoDC$ in RF WPT.

The above results show that, for a system employing low resistive load, employing high-PAPR signals to optimize RF WPT over a flat-fading channel, as suggested in  \cite{1G_Mobile_Power_Networks,reference,Clerckx_rectenna}, may not be suitable while employing a digital radio and thence for SWIPT too. Let us identify the probable cause for this result based on our simulation results in Fig.~\ref{fig:DAC_efficiency}. If we compare the plots for $A = \frac{1}{\sqrt{4}}$ and  $A = \frac{1}{4}$ from Fig.~\ref{fig:e2e_eff_b}, we clearly see that $A = \frac{1}{4}$ yields much lesser $\etaDCtoDC$ than $A = \frac{1}{\sqrt{4}}$ at all $G$ values even though we can observe in Fig.~\ref{fig:DAC_efficiency_b} that the PAPR of the RF waveform is higher with $A = \frac{1}{4}$ than with $A = \frac{1}{\sqrt{4}}$. If we now compare the average RF power of the two cases from Fig.~\ref{fig:DAC_efficiency_a}, the explanation seems obvious: It is much lower with $A = \frac{1}{4}$ than with $A = \frac{1}{\sqrt{4}}$. This implies that, while employing a digital radio, the factor that dominates the end-to-end efﬁciency performance of RF WPT over a ﬂat-fading channel is actually the average RF power that is transmitted and not the PAPR of the RF waveform. Moreover, it is evident from Fig.~\ref{fig:e2e_eff_c} that the maximum $\etaDCtoDC$ is achieved with $A=1$, when the average RF power is maximal as seen in Fig.~\ref{fig:DAC_efficiency_a}. This further validates our finding. This is in general true for any value of $N$.

We can now present a general analysis of the variation of $\etaDCtoDC$ w.r.t.\ $G$, which holds true for all the four cases. For $G$ between $40$~dB and $51$~dB, $\etaDCtoDC$ (in decibels) rises linearly for all the test waveforms. This region represents the combined linear amplification region of $\pai$ and $\pae$. For $G>51$~dB, the $\pai$ still operates in the linear amplification region, but the output RF power of the USRP gradually drives the $\pae$ into saturation, thence saturating $\etaDCtoDC$.

A remarkable observation is that the QPSK reference signal outperforms all the multisine signals at all $G$ values. A QPSK baseband signal has a consistent average RF power of one, as per (5) and (17), before the $\pai$ stage. Moreover, its average RF power is greater than that of any multisine with $N \ge 2$, as shown in Fig.~\ref{fig:DAC_efficiency_a}, for $A \le 1$. Consequently, the average radiated RF power after amplification by the PAs is also higher for the QPSK signal than for any multisine with $N \ge 2$. This reaffirms our hypothesis that it is the average radiated RF power of the waveform, and not its PAPR, that dominates the end-to-end efficiency of RF WPT when employing a digital radio transmitter over a flat-fading channel.

Interestingly, a QPSK baseband signal and a single-tone complex sinusoid signal with $A=1$ have the same baseband average power, and thus are expected to provide the same average radiated RF power, and consequently similar $\etaDCtoDC$. The inferior performance of $N=1$ sinusoid signal in Fig.~\ref{fig:e2e_eff} can be attributed to the practical limitations of a USRP. Due to digital pre-processing, its DAC appears to be truncating real (and imaginary) components of a complex sinusoidal signal whose magnitude is closer to one. Thus, the QPSK signal is unaffected by this limitation of the DAC. Consequently, $\PRFout$ for the the reference QPSK signal is higher than for $N=1$ case, which explains the difference in the resultant $\etaDCtoDC$.

\begin{table}[t]
\caption{Operational parameters for the experiments.}
\begin{center}
\begin{tabular}{ | m{3cm} | m{4.7cm}| } 
\hline
\textbf{Parameter} &  \textbf{Details} \\ 
\hline
Frequency band & $863-873$~MHz (unlicensed) \\ 
\hline
Carrier frequency & $\fc=868$~MHz \\ 
\hline
Sampling rate & $40$~MHz \\ 
\hline
Baseband signals & QPSK, 8-PSK, 8-QAM, 16-QAM\\ 
\hline
Bit rate & 1~Mbps, 20~Mbps, 40~Mbps\\ 
\hline
USRP gain setting & $G \in [40, 57]$~dB\\
\hline
\end{tabular}
\label{table:3}
\end{center}
\vspace{-12pt}
\end{table}

\begin{figure*}[t]
	\subfloat[End-to-end efficiency of information signals for varying USRP gain settings]{\includegraphics[width=0.32\textwidth]{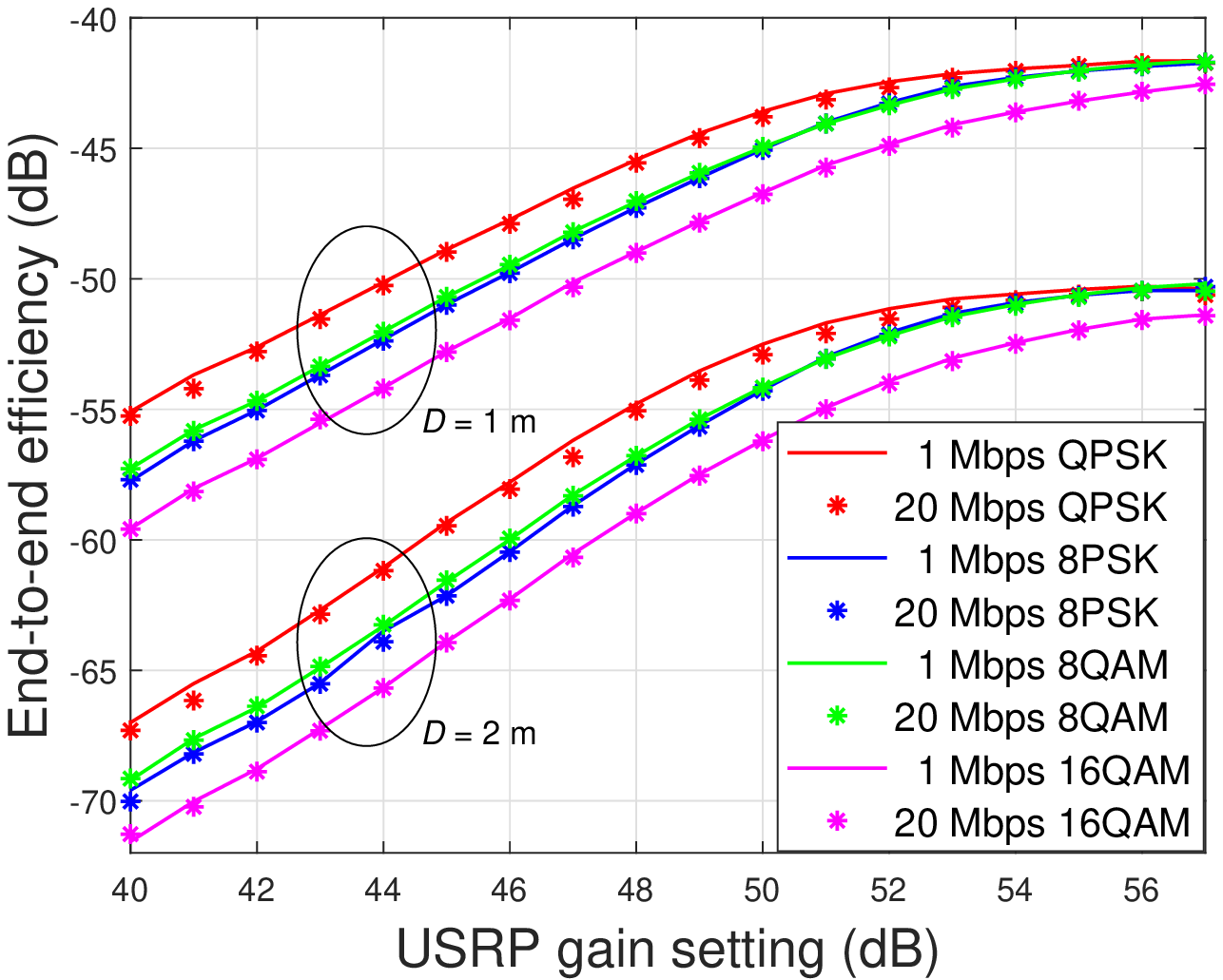}\label{fig:e2e_eff_mod_a}} \hspace{5pt}
	\subfloat[Transmitter and Receiver efficiency of information signals for varying USRP gain settings]{\includegraphics[,width=0.32\textwidth]{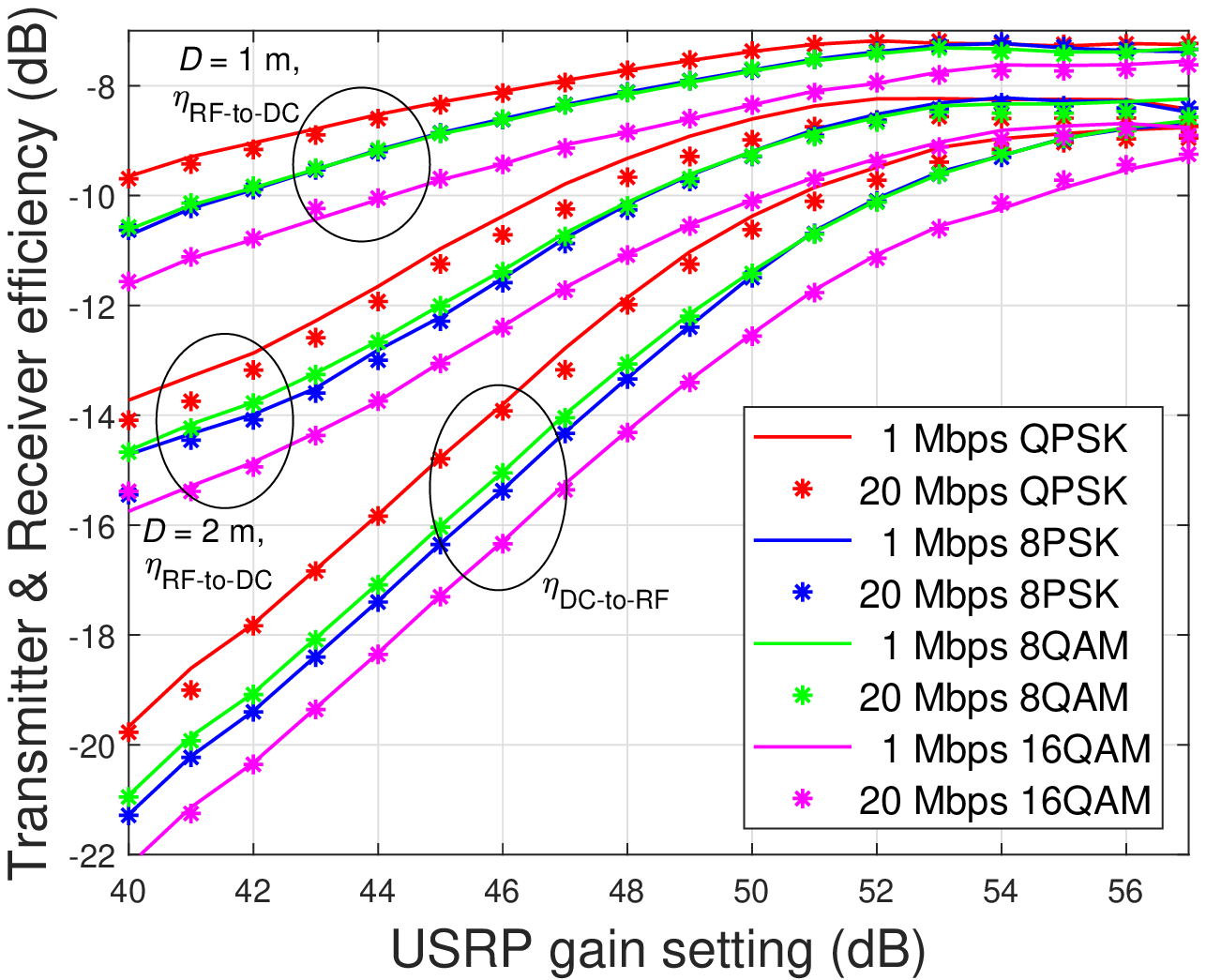}\label{fig:e2e_eff_mod_b}} \hspace{5pt}
    \subfloat[Channel efficiency of information signals for varying USRP gain settings]{\includegraphics[width=0.32\textwidth]{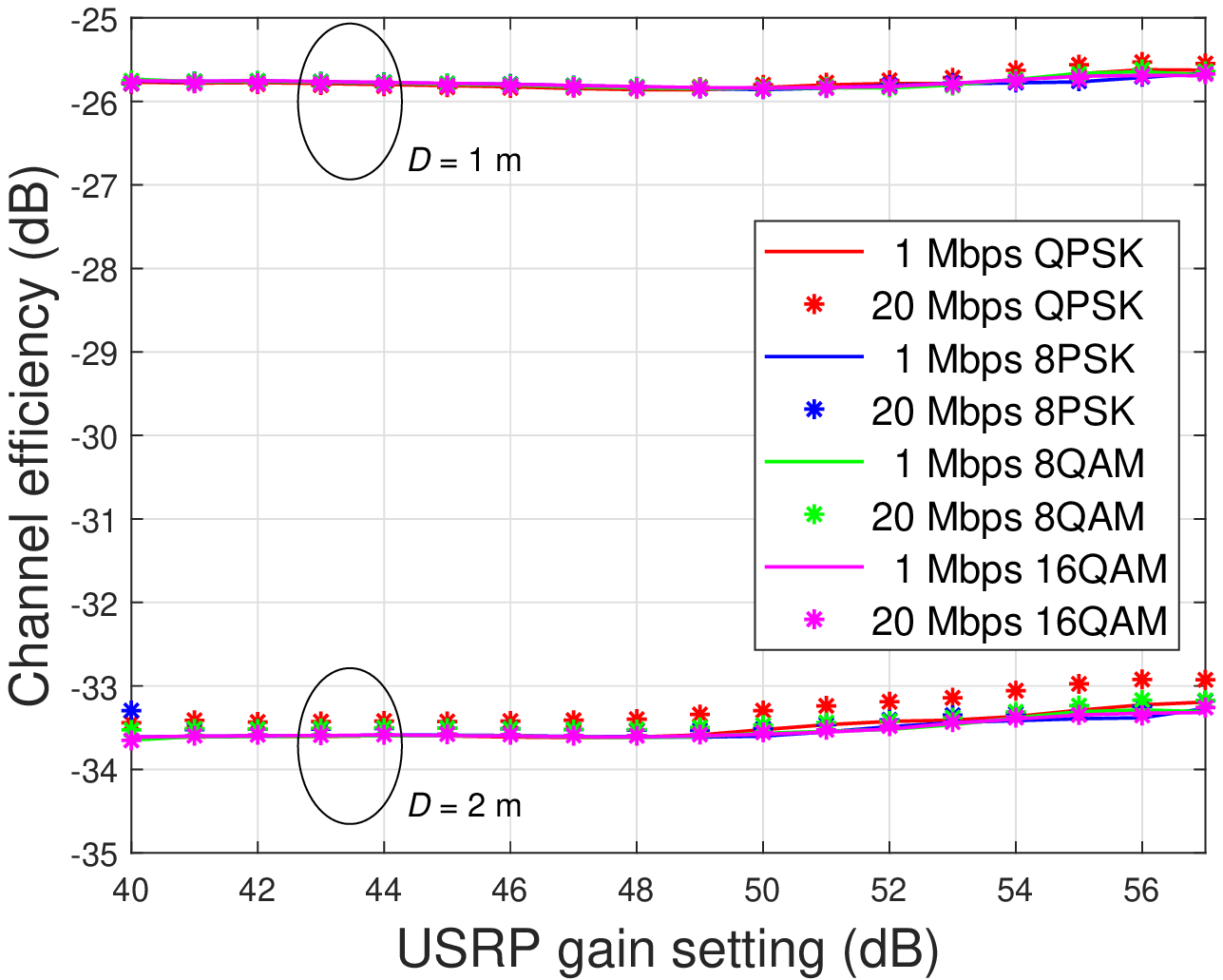}\label{fig:e2e_eff_mod_c}}
	\caption{\label{fig:e2e_eff_mod}Experimental observations for the end-to-end efﬁciency, transmitter efﬁciency and channel in RF WPT in terms of $G$ for QPSK, 8-PSK, 8-QAM and 16-QAM information signals at different bit rates. The corresponding receiver efficiency is computed using \eqref{eq:etaDCtoDC}. 
The efficiencies were evaluated for $D=1$~m and $2$~m, in the absence of a reflector.
	}
\end{figure*}

All the above observations are in contrast with those presented in literature \cite{bruno_arxiv}. The explanation for the disparity is as follows. The end-to-end efficiency of RF WPT is affected by numerous parameters such as the average power and PAPR of the waveform, CSI conditions, waveform shape, practical limitations of the SDR hardware, the energy harvester circuit and the load.  While we focus on $\etaDCtoDC$, the work in \cite{bruno_arxiv} focuses on $\etaRFtoRF$ and $\etaRFtoDC$. Consequently, we have the $\PDCin$ fixed in our work, while \cite{bruno_arxiv} keeps the RF transmit power the same. Thus, the radiated power levels are very different in the two works. Moreover, the researchers in \cite{bruno_arxiv} use a fabricated rectifier circuit, while we utilize an off-the-shelf evaluation board for an EH receiver. Apart from these, the other important difference is the load. The results in \cite{bruno_arxiv} are based on a resistive load of $10$~k$\Omega$ and $12$~k$\Omega$, while we employ $R=286$~$\Omega$. The impact of variation in the resistive load on $\etaRFtoDC$ of RF WPT is presented in \cite{Bruno_contrast_Robust_Wireless_Power_Receiver_for_Multi-Tone_Waveforms}, where the observations for $\etaRFtoDC$ concur with our observations for multisine efficiencies with low resistive load. 

The observation that a QPSK signal provides even better $\etaDCtoDC$ than multisine signals pits them as a potential candidate waveform for SWIPT. Based on these results, we continue our further analysis of RF WPT considering only baseband signals that are already being used for information transfer.

\subsection{End-to-End Efficiency of Communication Signals versus $G$}\label{sec:e2eeff_info_vs_gain}

In what follows, we concentrate on QPSK,  8-PSK, \mbox{8-QAM} and 16-QAM signals and determine their performance in RF WPT/SWIPT. In particular, the constellation points of  8-QAM are $\pm j/\sqrt{2}$, $\pm 1/\sqrt{2}$, and $\pm 1/\sqrt{2} \pm j/\sqrt{2}$. These digital baseband signals are defined such that they avoid any clipping by the DAC. Thus, the average power of the digital baseband QPSK, 8-PSK, 8-QAM and 16-QAM signals are $1$, $1$, $0.75$, and $0.56$, respectively. The operational parameters are presented in Table~\ref{table:3}. Given a $10$~MHz channel bandwidth, we limit the bit rate of the baseband signals to $20$~Mbps.

The results for the $\etaDCtoDC$ performance of $1$~Mbps and $20$~Mbps bit rate communication signals for varying $G$ are depicted in Fig.~\ref{fig:e2e_eff_mod_a}. We observe that for a given bit rate, say $1$~Mbps, $\etaDCtoDC$ reduces as the modulation order increases. So, a QPSK-modulated signal  has better $\etaDCtoDC$ than 8-PSK and 8-QAM signals, which in turn fare better than a 16-QAM signal. In addition, we observe that for a given baseband modulation, $\etaDCtoDC$ is invariant of the bit rate. In general, $\etaDCtoDC$ increases linearly with  $G$ in the combined linear region of amplification of the two PAs, up to $G=51$~dB, and then $\etaDCtoDC$ begins to saturate as the $\pae$ operates in the saturation region. Moreover, the observations are similar for both propagation distances.

\subsection{Transmitter, Channel and Receiver Efficiency of Communication Signals versus $G$}\label{sec:txchreff}

Let us now examine the end-to-end efficiency of communication signals in detail by evaluating $\etaDCtoRF$, $\etaRFtoRF$ and $\etaRFtoDC$ separately. First, we determine $\etaRFtoDC$ by employing the VST as a receiver as explained Section~\ref{sec:test-bed-tx_eff}, with the test-bed depicted in Fig.~\ref{fig:test-bed_b}. The results are shown in Fig.~\ref{fig:e2e_eff_mod_b}. The maximum attainable $\etaDCtoRF$ is about $-8.7$~dB ($13.5\%$). The remaining observations are similar to those in Section~\ref{sec:e2eeff_info_vs_gain}. 

Afterwards, we evaluate $\etaRFtoRF$ (path loss) in RF WPT. The test-bed for evaluating $\etaRFtoRF$ is presented in Fig.~\ref{fig:test-bed_c}. The results are presented in Fig.~\ref{fig:e2e_eff_mod_c}. We observe that all the signals experience similar path loss even at $20$~Mbps bit rate, which corresponds to $10$~MHz, $6.67$~MHz and $5$~MHz bandwidth for undistorted QPSK, 8-PSK/8-QAM and \mbox{16-QAM} signals, respectively. This observation concurs with the one in Section~\ref{subsection:channel_sounding} about the channel being almost flat-fading. However, once the $\pae$ starts operating in the saturation mode ($G>51$~dB), we observe deviations in $\etaRFtoRF$, significantly so for the $20$~Mbps QPSK signal. This arises because of the non-linear distortions that lead to spectral regrowth and subject the signals to a frequency-selective channel.

\begin{figure*}[t]
\centering
\begin{minipage}[b]{.32\textwidth}
 \includegraphics[width=\textwidth]{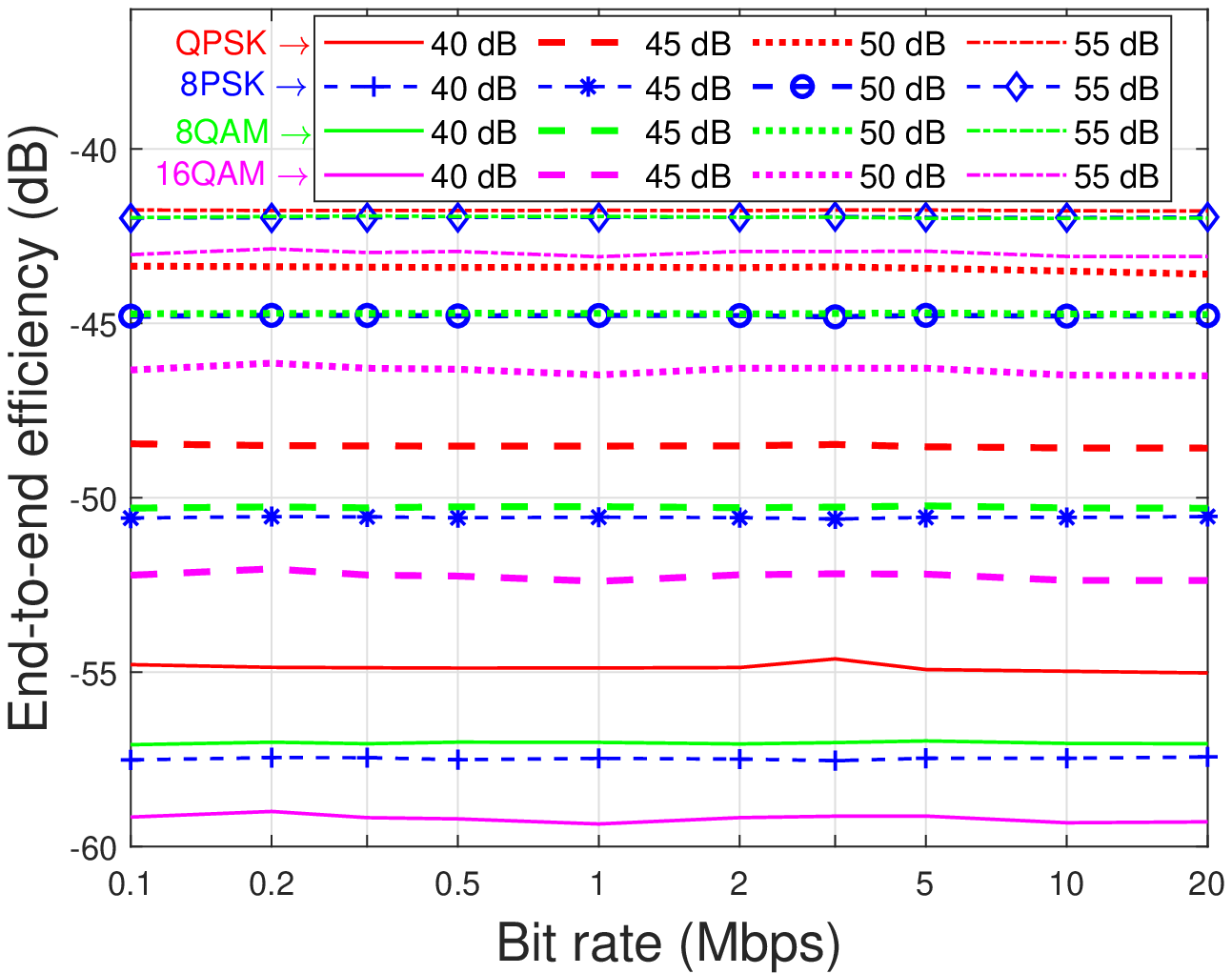}
\caption{End-to-end efﬁciency of information signals for varying bit rates.} \label{fig:e2e_eff_rate}
\end{minipage}\hfill
\begin{minipage}[b]{.32\textwidth}
\includegraphics[width=\textwidth]{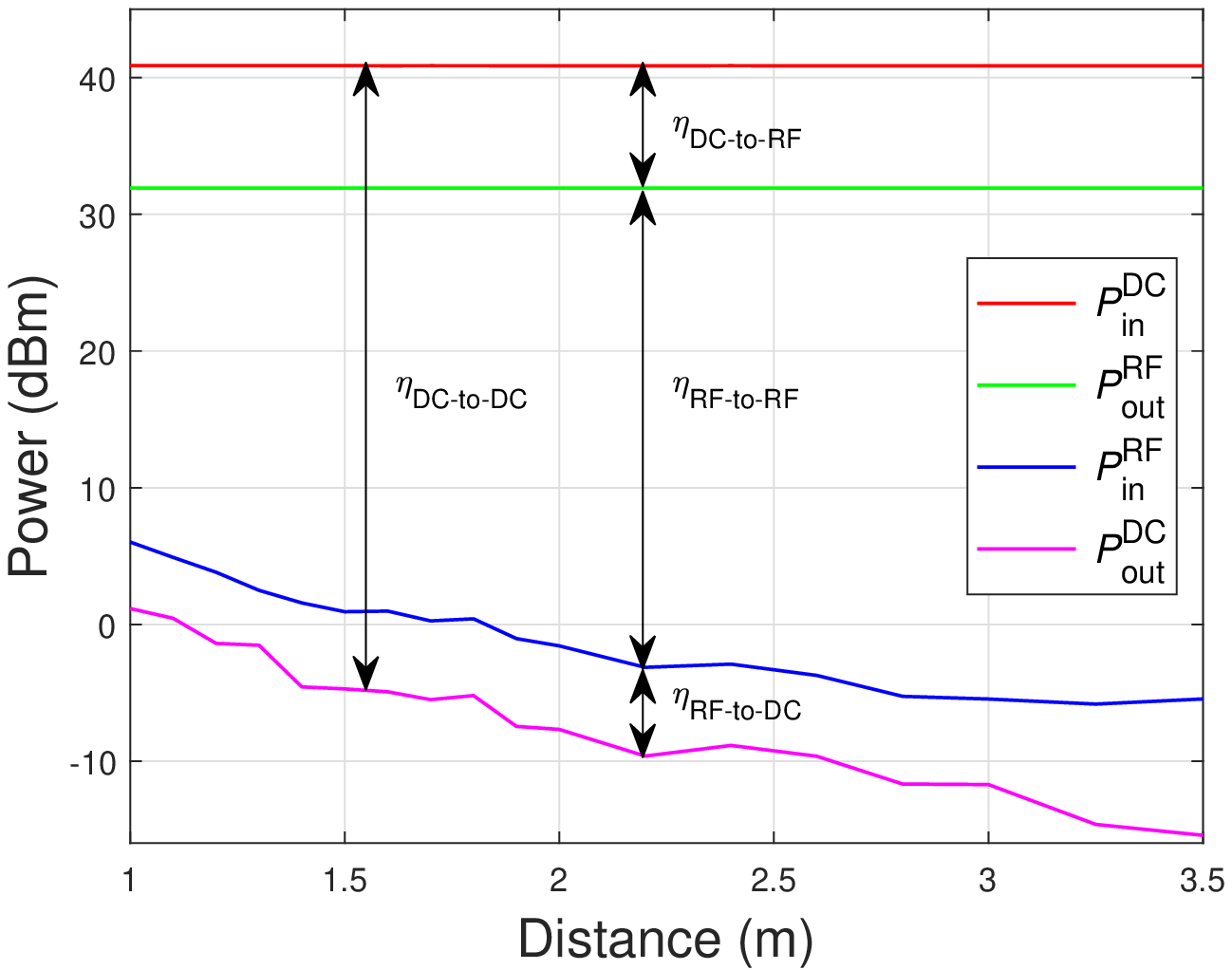}
\caption{Evaluation of the RF WPT performance of a QPSK-modulated signal for varying $D$, in terms of various efficiencies.}\label{fig:e2e_eff_dist}
\end{minipage}\hfill
\begin{minipage}[b]{.32\textwidth}
\includegraphics[width=\textwidth]{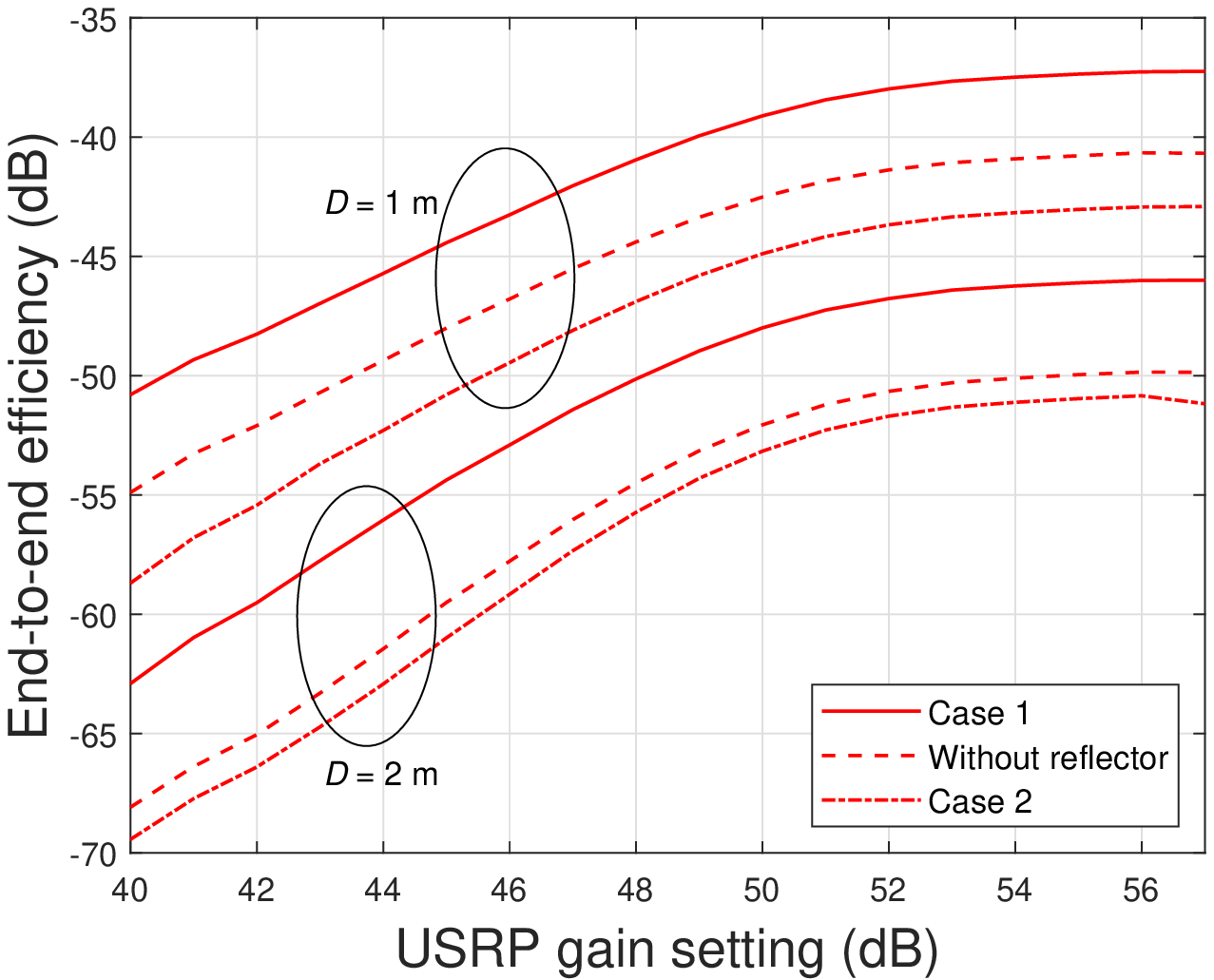}
\caption{End-to-end efﬁciency of a QPSK modulated signal for varying channel conditions.}\label{fig:e2e_eff_dist_reflector}
\end{minipage}
\end{figure*}

Based on the available measurement data, we compute the corresponding $\etaRFtoDC$ using \eqref{eq:etaDCtoDC} and the results are as shown in Fig.~\ref{fig:e2e_eff_mod_b}. The primary observation, which has been missed in the previous literature, is that $\etaRFtoDC<\etaDCtoRF$. For $D=1$~m, the maximum $\etaRFtoDC$ is around $-7$~dB (about $20\%$), when $\PRFin\approx 6$~dBm\footnote{In comparison, the rectennas in \cite{Clerckx_rectenna}  and \cite{Rectenna_collado} offer maximum efficiency of about $12\%$ and $15\%$ respectively, for $\PRFin\approx -20$~dBm at $2.4$~GHz.}. A remarkable observation in Fig.~\ref{fig:e2e_eff_mod_b} is that, in the linear amplification region, the slopes of the curves for $D=1$~m and $2$~m are dissimilar. This implies that $\etaRFtoDC$ of the employed harvester increases rapidly at lower $\PRFin$ levels. The remaining observations are similar to those in Section~\ref{sec:e2eeff_info_vs_gain}. The receiver efficiency is similar to the observations in \cite{reference} for the given $\PRFin$ levels. However, due to the high path losses, characterization of $\etaRFtoDC$ at higher $\PRFin$ is not feasible in our RF WPT experiments and needs to be studied separately.

Another significant observation is that although the digital samples of QPSK and 8-PSK waveforms have the same average power,  the former has superior $\etaDCtoDC$ (resp.\ $\etaDCtoRF$, $\etaRFtoDC$) performance for most $G$ values. Conversely, \mbox{8-PSK} and 8-QAM waveforms have similar $\etaDCtoDC$ (resp.\ $\etaDCtoRF$, $\etaRFtoDC$) performance at all $G$ values, even though the digital samples of the latter have lower average power. These observations can again be attributed to the practical limitations of the USRP in handling the real (and imaginary) values of a complex sample, whose magnitude is close to one. 

Overall, from the foregoing experiments, we observe that a QPSK modulated signal provides the best $\etaDCtoDC$ performance at all bit rate and $G$ values, and thus it seems to be the most suitable choice from the RF SWIPT perspective so far.

\subsection{End-to-End Efficiency of Communication Signals versus Transmission Bit Rate}\label{sec:e2e_eff_vs_rate}

A common observation in Fig.~\ref{fig:e2e_eff_mod} is that all the efficiencies are invariant of the transmission bit rate. Thus, let us explore further the dependence of $\etaDCtoDC$ to the information signals' bit rate. The setup employed is the one in  Section~\ref{sec:e2e_eff}, and the operational parameters are as presented in Table~\ref{table:3}, with the notable changes being in the values of the bit rate ($100$~kbps, $200$~kbps, $320$~kbps, $500$~kbps, $1$~Mbps, $2$~Mbps,  $3.2$~Mbps,  $5$~Mbps, $10$~Mbps and $20$~Mbps) and $G$ ($40$~dB, $45$~dB, $50$~dB and $55$~dB). The results are shown in Fig.~\ref{fig:e2e_eff_rate}. 

The primary observation is that the end-to-end efficiency of all the information signals is indeed fairly independent of their bit rate. It must be noted that we examine the {\em transmit} bit rate here. The harvested energy, however, can be strongly dependent on the {\em receive} data rate \cite{Yoo_A_Dual-Circular-Polarized_Endoscopic_Antenna_With_Wideband_Characteristics_and_Wireless_Biotelemetric_Link_Characterization}. The other observations are similar to those in Fig.~\ref{fig:e2e_eff_mod_a}. These observations lead us to conclude that for SWIPT, the choice of bit rate (for digitally modulated transmission) would not be limited by $\etaDCtoDC$.

\subsection{Efficiency versus Transmitter--Receiver Distance and Varying Channel Conditions}
Next, we observe the impact of varying $D$ on the various efficiencies of an RF WPT system at once.  We measured $\PDCin, \PRFout, \PRFin$ and $\PDCout$ for a $1$~Mbps QPSK waveform by varying the position of the receiver from $D=1$~m to $3.5$~m, and the results are depicted in Fig.~\ref{fig:e2e_eff_dist}. In accordance with the observation in \cite{Yoo_A_Metamaterial-Coupled_Wireless_Power_Transfer_System_Based_on_Cubic_High-Dielectric_Resonators}, we notice that $\etaRFtoDC$ is dependent on $D$, and hence on $\PRFin$. As  $\PRFin$ reduces, $\etaRFtoDC$ decreases as well. Fig.~\ref{fig:e2e_eff_dist} provides us with a view of the viability of RF WPT. We observe that for a distance of $3$~m, our test-bed is able to harvest about $0.1$~mW of DC power, while for $1$~m separation the harvested power is higher than $1$~mW. Figure~\ref{fig:e2e_eff_dist} also provides us with a visual depiction of the relation between the three efficiencies: $\etaRFtoRF \ll \etaDCtoRF < \etaRFtoDC$.

Let us now observe the impact of a reflecting surface on $\etaDCtoDC$ performance in RF WPT for all the six cases shown in Fig.~\ref{fig:chamber_etaDCtoDC} and discussed in Section~\ref{subsection:channel_sounding}. We employ a $1$~Mbps QPSK-modulated signal for this experiment. In  Fig.~\ref{fig:e2e_eff_dist_reflector}, we observe a significant improvement of about $3-4$~dB in $\etaDCtoDC$ for Case 1 and a deterioration of about $2-3$~dB is seen in Case 2 when $D=1$~m. For $D=2$~m, an improvement of around $4-5$~dB is observed in Case 1 with respect to the reference case, while the diminution in Case 2 is approximately $1$~dB. The observations suggest that a properly positioned reflector can effectively double $\PDCout$ in RF WPT. In a dynamic real-world scenario, multiple such reflecting surfaces, both fixed and mobile, may be present between the transmitter and the energy harvesting receiver, and thus their impact on $\etaDCtoDC$ must be taken into account while designing the optimal waveforms for RF WPT/SWIPT.

\begin{figure*}[ht]
	\centering
	\subfloat[EVM of information signals for varying USRP gain settings]{\includegraphics[width=0.45\textwidth]{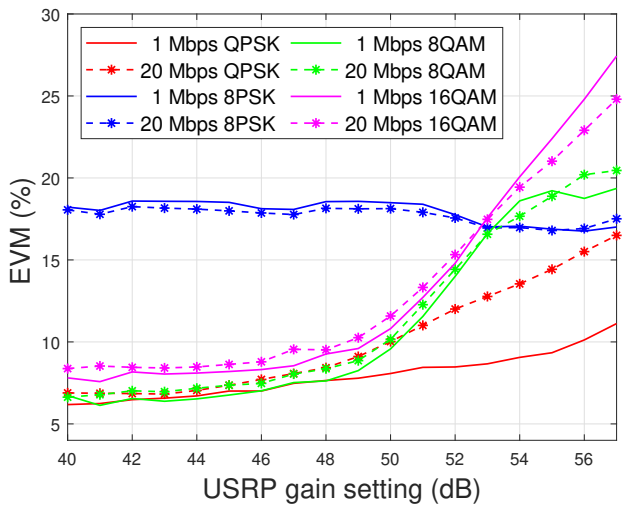}\label{fig:evm_gain}}	
	\hspace{25pt}
    \subfloat[EVM of information signals for varying  bit rates]{\includegraphics[width=0.45\textwidth]{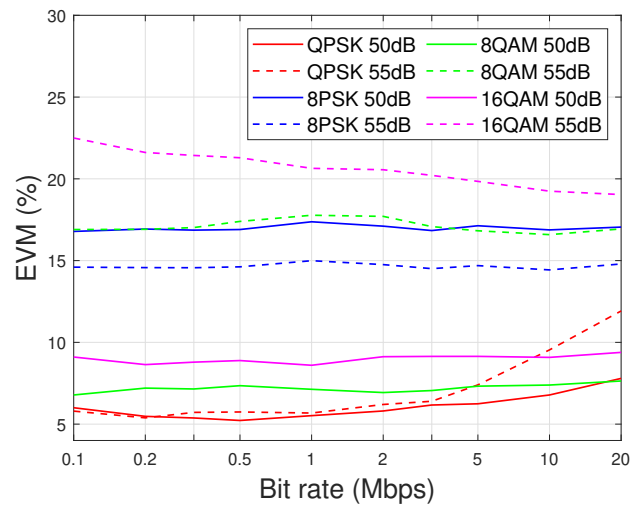}\label{fig:evm_rate}}
	\caption{\label{fig:evm} The measured EVM of the communication signals at different bit rates and $G$ values. The complex baseband symbols were captured by the VST, and compensated for the CFO, before evaluating the EVM. Overall, the QPSK signal yields the lowest EVM for any given bit rate and G.}
\end{figure*}

\subsection{Performance Analysis of PSK and QAM Modulations}\label{sec:experimental_EVM_analysis}

Apart from the experiments, utilizing a digital radio for WPT is worthwhile mainly if we target SWIPT. Consequently, we ought to determine the impact of varying $G$ and bit rates on the quality of the transmission signals to find the modulation that would be suitable for information transfer. To do so, we have measured the EVM of  communication signals. The  operational parameters are those noted in Table~\ref{table:3}, unless mentioned otherwise. For each of these measurements, we captured $7.2$ million samples (6 repetitions $\times$ 50 batches/repetition $\times$ 24000 samples/batch) of each transmitted signal and computed the EVM. It must be noted that the oscillator at the USRP transmitter is not extremely stable resulting in carrier frequency offset (CFO), which needs to be compensated before computing the EVM. The justification for the observations in Fig.~\ref{fig:evm} is based on the analysis of the constellation of the captured samples (not shown herein). 

\subsubsection{EVM versus $G$}
The EVM response of $1$~Mbps and $20$~Mbps information signals for varying $G$ is shown in Fig.~\ref{fig:evm_gain}. To begin, we observe that, except for the QPSK-modulated signals, all the other information signals have similar EVM irrespective of the bit rates. Let us present a brief explanation for these EVM curves.

In the case of QPSK signals, the effect of CFO is observed to be more prominent in the samples for $20$~Mbps signal, such that it could not be completely compensated, thus resulting in higher EVM. Overall, for the $1$~Mbps QPSK signal, the EVM is fairly low for $G \le 50$~dB with a slight increase after that, while for the $20$~Mbps QPSK signal, the EVM stays consistently low till $G = 45$~dB and increases rapidly thereafter. The higher EVM for the $20$~Mbps QPSK signal can attributed to the increased wideband noise that creeps in at higher $G$ values.

Coming to 8-PSK signals, we observe in Fig.~\ref{fig:evm_gain} that the EVM remains high until  $G = 51$~dB and reduces marginally thereafter. The constellation plots reveal that the reason for the high EVM is the distinct rotation of different constellation points. We observe that for $G < 51$~dB, symbols within a quadrant tend to merge due to CFO. This impact of the CFO starts reducing gradually thereafter due to which the symbols begin to realign with the standard 8-PSK constellation, and the EVM decreases.

Next, for 8-QAM signals, we observe that the EVM initially increases gradually with $G$ up to $50$~dB, and then shows a sharp increase. As $G$ increases, we notice that the \mbox{8-QAM} constellation begins to shift towards an 8-PSK constellation. This occurs due to the distinct power levels \mbox{8-QAM} symbols: $0.5$ (when $\pm 1/\sqrt{2}$ or $\pm 1j/\sqrt{2}$), while one for others. Consequently, before the $\pai$ stage, the RF signals corresponding to symbols with higher average power will have higher amplitude. So long as the RF signals are linearly amplified ($G<51$~dB), the EVM will be low. Thereafter, the RF signals corresponding to $\pm 1/\sqrt{2}\pm 1j/\sqrt{2}$ already saturate the $\pai$ due to their higher amplitudes, while the others continue to be linearly amplified. At the baseband, this translates to a gradual shifting of the 8-QAM constellation towards the 8-PSK one. Nonetheless, for $G>55$~dB, the RF signals corresponding to all the 8-QAM symbols eventually saturate the $\pai$. This explains the increase in EVM for $G$ between $50$--$55$~dB and the subsequent flattening of the curve as seen in Fig.~\ref{fig:evm_gain}. 

Finally, the case of 16-QAM signals is similar to 8-QAM, as can be observed from Figs.~\ref{fig:evm_gain}. The corresponding explanation is same as well. If we now focus on the overall plot, we observe that for any $G$, the QPSK-modulated signal yields the lowest EVM for both the transmission bit rates.

\subsubsection{EVM versus Bit Rate}
Next, we evaluated the EVM performance of the information signals by varying their bit rate for some $G$ values. The observations of the previous experiment  help us deduce that the EVM of the information signals does not vary much for $G < 50$~dB, and so we chose  $G=50$~dB and $55$~dB for these evaluations. We observe in Fig.~\ref{fig:evm_rate} that the EVM for 8-PSK, 8-QAM and 16-QAM signals are invariant of the transmission bit rate. With a QPSK signal,  we observe that the EVM is fairly steady up to $2$~Mbps bit rate, and then increases gradually in the linear amplification region, while rapidly in the saturation region of operation. The constellation analysis of QPSK-modulated signal reveals that the gradual increase of EVM beyond $2$~Mbps bit rate is due to the impact of CFO which occurs over clusters of samples, while the rapid increase in the saturation region of operation can be attributed to the wideband noise that creeps in.

Overall, we observe that even with the EVM degrading at higher data rates, QPSK-modulated signals are still the best choice among the selected communication signals. The CFO issues would be resolved by utilizing a more stable oscillator at the transmitter. The observations from the above three experiments lead us to propose that, when employing a digital radio transmitter, QPSK-modulated signals are best-suited for SWIPT, both in terms of their end-to-end efficiency of RF WPT and the transmission signal quality.

%%%%%%%%%%%%%%%%%%%%%%%%%%%%%%%%%%%%%%%%%%%%%%%%%%%%%%%%%%%%%%%%%%%%%%%%%%%%%%%%%%%%%%%%%%%%%%%%%%%%%%%%%%%%%%%%%%%%%%%%%%%%%%%%%%%%%%%%%%%%%%%%%%%%%%%%%%%%%%%%
\section{Conclusion and Future Work}\label{sec:conclusions}

This paper theoretically analyzes the implications of signal processing in a digital radio on multisine waveforms and experimentally determines the modulated waveforms suitable for RF WPT and SWIPT. We explained the manner in which the DAC and internal PA of a digital radio curtail the average radiated RF power of multisine waveforms. Furthermore, we developed an SDR-based test-bed to evaluate the end-to-end efficiency of RF WPT over a flat-fading channel and observed that a QPSK signal outperforms the multisine signals, when the resistive load is of the order of a few hundred ohms. Moreover, the end-to-end efficiency of communication signals is independent of their transmission bit rate, and it may be significantly enhanced by constructive multipath propagation. We demonstrated that, in addition to over-the-air attenuation, the low DC-to-RF efficiency at the transmitter is the other prominent reason for the low end-to-end efficiency in RF WPT. Overall, with the current setup, a QPSK waveform yields the highest end-to-end efficiency and the lowest EVM among PSK and QAM signals, and thus is most favorable for SWIPT.

While the eventual goal of our research work is to develop a prototype for SWIPT, we intend on addressing several interesting research problems in future works by leveraging the test-bed and results from the current work. These include experimentally studying the impact of varying resistive loads on the $\etaDCtoDC$ performance, determining waveforms that optimize the $\etaDCtoDC$  of RF WPT as well as evaluating the $\etaDCtoDC$ performance under regulatory power constraints. We also intend on replacing the current harvester with our own fabricated rectifier and examining the differences. Furthermore, our test-bed enables us to pursue rectifier modeling.

%%%%%%%%%%%%%%%%%%%%%%%%%%%%%%%%%%%%%%%%%%%%%%%%%%%%%%%%%%%%%%%%%%%%%%%%%%%%%%%%%%%%%%%%%%%%%%%%%%%%%%%%%%%%%%%%%%%%%%%%%%%%%%%%%%%%%%%%%%%%%%%%%%%%%%%%%%%%%%%%

\bibliographystyle{IEEEtran}
\bibliography{IEEEabrv,TMTT-2020-09-1035_Citation1,TMTT-2020-09-1035_Citation2,TMTT-2020-09-1035_Citation3}

\begin{thebibliography}{10}
\providecommand{\url}[1]{#1}
\csname url@samestyle\endcsname
\providecommand{\newblock}{\relax}
\providecommand{\bibinfo}[2]{#2}
\providecommand{\BIBentrySTDinterwordspacing}{\spaceskip=0pt\relax}
\providecommand{\BIBentryALTinterwordstretchfactor}{4}
\providecommand{\BIBentryALTinterwordspacing}{\spaceskip=\fontdimen2\font plus
\BIBentryALTinterwordstretchfactor\fontdimen3\font minus
  \fontdimen4\font\relax}
\providecommand{\BIBforeignlanguage}[2]{{%
\expandafter\ifx\csname l@#1\endcsname\relax
\typeout{** WARNING: IEEEtran.bst: No hyphenation pattern has been}%
\typeout{** loaded for the language `#1'. Using the pattern for}%
\typeout{** the default language instead.}%
\else
\language=\csname l@#1\endcsname
\fi
#2}}
\providecommand{\BIBdecl}{\relax}
\BIBdecl

\bibitem{Nachiket-IMS}
N.~{Ayir}, M.~F. {Trujillo Fierro}, T.~{Riihonen}, and M.~{All\'en},
  ``Experimenting waveforms and efficiency in {RF} power transfer,'' in
  \emph{Proc. IEEE MTT-S International Microwave Symposium (IMS)}, Jun. 2019.

\bibitem{1G_Mobile_Power_Networks}
B.~{Clerckx}, A.~{Costanzo}, A.~{Georgiadis}, and N.~B. {Carvalho}, ``Toward
  {1G} mobile power networks: {RF}, signal, and system designs to make smart
  objects autonomous,'' \emph{IEEE Microwave Magazine}, vol.~19, no.~6, pp.
  69--82, Sep. 2018.

\bibitem{Magnetic_resonance_Kurs}
\BIBentryALTinterwordspacing
A.~Kurs, A.~Karalis, R.~Moffatt, J.~D. Joannopoulos, P.~Fisher, and M.~Solja{\v
  c}i{\'c}, ``Wireless power transfer via strongly coupled magnetic
  resonances,'' \emph{Science}, vol. 317, no. 5834, pp. 83--86, 2007. [Online].
  Available: \url{https://science.sciencemag.org/content/317/5834/83}
\BIBentrySTDinterwordspacing

\bibitem{Inductive_WPT_specs}
R.~{Tseng}, B.~{von Novak}, S.~{Shevde}, and K.~A. {Grajski}, ``Introduction to
  the alliance for wireless power loosely-coupled wireless power transfer
  system specification version 1.0,'' in \emph{Proc. IEEE Wireless Power
  Transfer Conference (WPTC)}, May 2013.

\bibitem{Qi_standard}
{Wireless Power Consortium (WPC)}. Qi specifications. [Online]. Available:
  “https://www.wirelesspowerconsortium.com/knowledge-base/specifications/",
  Accessed on 27 Aug. 2020.

\bibitem{AirFuel}
{AirFuel Alliance}. {AirFuel Resonant}. [Online]. Available:
  “https://airfuel.org/wireless-power/electromagnetic-coupling/", Accessed on
  27 Aug. 2020.

\bibitem{A_review_of_WPT}
M.~{Rayes}, G.~{Nagib}, and W.~{Abdelaal}, ``A review on wireless power
  transfer,'' \emph{International Journal of Engineering Trends and Technology
  (IJETT)}, vol.~40, no.~5, pp. 272--280, Oct. 2016.

\bibitem{comparison_induction_resonance_WPC}
{Wireless Power Consortium}. Magnetic resonance and magnetic induction.
  [Online]. Available: “https://www.wirelesspowerconsorti
  um.com/data/downloadables/1/2/4/6/magnetic-resonance-or-magnetic-induction.pdf",
  Accessed on 27 Aug. 2020.

\bibitem{comparison_induction_resonance_DigiKey}
{DigiKey}. Inductive versus resonant wireless charging: A truce may be a
  designer’s best choice. [Online]. Available:
  “https://www.digikey.fi/en/articles/inductive-versus-resonant-wireless-charging",
  Accessed on 27 Aug. 2020.

\bibitem{comparison_induction_resonance_RFWPT_Ansys}
{Ansys}, ``Wireless charging technologies: Magnetic resonance vs. magnetic
  induction vs. {RF} harvesting,'' [Online]. Available:
  “https://www.ansys.com/blog/ces-wireless-charging-magnetic-resonance-induction-rf-harvesting/",
  Accessed on 27 Aug. 2020.

\bibitem{Koomey}
J.~Koomey, S.~Berard, M.~Sanchez, and H.~Wong, ``Implications of historical
  trends in the electrical efficiency of computing,'' \emph{IEEE Annals of the
  History of Computing}, vol.~33, no.~3, pp. 46--54, Mar. 2011.

\bibitem{SWIPT-Krikidis}
I.~{Krikidis}, S.~{Timotheou}, S.~{Nikolaou}, G.~{Zheng}, D.~W.~K. {Ng}, and
  R.~{Schober}, ``Simultaneous wireless information and power transfer in
  modern communication systems,'' \emph{IEEE Communications Magazine}, vol.~52,
  no.~11, pp. 104--110, Nov. 2014.

\bibitem{reference}
B.~{Clerckx}, R.~{Zhang}, R.~{Schober}, D.~W.~K. {Ng}, D.~I. {Kim}, and H.~V.
  {Poor}, ``Fundamentals of wireless information and power transfer: From {RF}
  energy harvester models to signal and system designs,'' \emph{IEEE Journal on
  Selected Areas in Communications}, vol.~37, no.~1, pp. 4--33, Jan. 2019.

\bibitem{Survey-Hu}
J.~{Hu}, K.~{Yang}, G.~{Wen}, and L.~{Hanzo}, ``Integrated data and energy
  communication network: A comprehensive survey,'' \emph{IEEE Communications
  Surveys and Tutorials}, vol.~20, no.~4, pp. 3169--3219, Fourthquarter 2018.

\bibitem{Survey-Tharindu}
T.~D. {Ponnimbaduge Perera}, D.~N.~K. {Jayakody}, S.~K. {Sharma},
  S.~{Chatzinotas}, and J.~{Li}, ``Simultaneous wireless information and power
  transfer ({SWIPT}): Recent advances and future challenges,'' \emph{IEEE
  Communications Surveys and Tutorials}, vol.~20, no.~1, pp. 264--302,
  Firstquarter 2018.

\bibitem{Survey_beamforming}
Y.~{Alsaba}, S.~K.~A. {Rahim}, and C.~Y. {Leow}, ``Beamforming in wireless
  energy harvesting communications systems: A survey,'' \emph{IEEE
  Communications Surveys and Tutorials}, vol.~20, no.~2, pp. 1329--1360,
  Secondquarter 2018.

\bibitem{Survey_efficiency}
C.~R. {Valenta} and G.~D. {Durgin}, ``Harvesting wireless power: Survey of
  energy-harvester conversion efficiency in far-field, wireless power transfer
  systems,'' \emph{IEEE Microwave Magazine}, vol.~15, no.~4, pp. 108--120, Jun.
  2014.

\bibitem{Survey_networks}
X.~{Lu}, P.~{Wang}, D.~{Niyato}, D.~I. {Kim}, and Z.~{Han}, ``Wireless networks
  with {RF} energy harvesting: A contemporary survey,'' \emph{IEEE
  Communications Surveys and Tutorials}, vol.~17, no.~2, pp. 757--789,
  Secondquarter 2015.

\bibitem{Magazine_Carvalho}
A.~J.~S. {Boaventura}, D.~{Belo}, R.~{Fernandes}, A.~{Collado},
  A.~{Georgiadis}, and N.~B. {Carvalho}, ``Boosting the efficiency:
  Unconventional waveform design for efficient wireless power transfer,''
  \emph{IEEE Microwave Magazine}, vol.~16, no.~3, pp. 87--96, Apr. 2015.

\bibitem{BC1}
B.~{Clerckx} and E.~{Bayguzina}, ``Waveform design for wireless power
  transfer,'' \emph{IEEE Transactions on Signal Processing}, vol.~64, no.~23,
  pp. 6313--6328, Dec. 2016.

\bibitem{Carvalho_diode_equation}
A.~J.~S. {Boaventura} and N.~B. {Carvalho}, ``Maximizing {DC} power in energy
  harvesting circuits using multisine excitation,'' in \emph{Proc. IEEE MTT-S
  International Microwave Symposium}, Jun. 2011.

\bibitem{Elena1}
E.~{Boshkovska}, D.~W.~K. {Ng}, N.~{Zlatanov}, and R.~{Schober}, ``Practical
  non-linear energy harvesting model and resource allocation for {SWIPT}
  systems,'' \emph{IEEE Communications Letters}, vol.~19, no.~12, pp.
  2082--2085, Dec. 2015.

\bibitem{Rui1}
M.~R.~V. {Moghadam}, Y.~{Zeng}, and R.~{Zhang}, ``Waveform optimization for
  radio-frequency wireless power transfer,'' in \emph{Proc. IEEE 18th
  International Workshop on Signal Processing Advances in Wireless
  Communications (SPAWC)}, Jul. 2017.

\bibitem{BC2}
Y.~{Huang} and B.~{Clerckx}, ``Waveform design for wireless power transfer with
  limited feedback,'' \emph{IEEE Transactions on Wireless Communications},
  vol.~17, no.~1, pp. 415--429, Jan. 2018.

\bibitem{Kim_imperfectCSI}
L.~{Cantos}, G.~{Sacarelo}, Y.~{Jeong}, and Y.~H. {Kim}, ``Performance of a
  waveform design for wireless power transfer with imperfect channel state
  information,'' in \emph{Proc. International Conference on Information and
  Communication Technology Convergence (ICTC)}, Oct. 2017.

\bibitem{Clerckx_transmit_diversity}
B.~{Clerckx} and J.~{Kim}, ``On the beneficial roles of fading and transmit
  diversity in wireless power transfer with nonlinear energy harvesting,''
  \emph{IEEE Transactions on Wireless Communications}, vol.~17, no.~11, pp.
  7731--7743, Nov. 2018.

\bibitem{Elena_Robust_Beamforming}
E.~{Boshkovska}, A.~{Koelpin}, D.~W.~K. {Ng}, N.~{Zlatanov}, and R.~{Schober},
  ``Robust beamforming for {SWIPT} systems with non-linear energy harvesting
  model,'' in \emph{Proc. IEEE 17th International Workshop on Signal Processing
  Advances in Wireless Communications (SPAWC)}, Jul. 2016.

\bibitem{Sun_Robust_Beamforming}
H.~{Sun}, F.~{Zhou}, and Z.~{Zhang}, ``Robust beamforming design in a {NOMA}
  cognitive radio network relying on {SWIPT},'' in \emph{Proc. IEEE
  International Conference on Communications (ICC)}, May 2018.

\bibitem{bruno_arxiv}
J.~{Kim}, B.~{Clerckx}, and P.~D. {Mitcheson}, ``Signal and system design for
  wireless power transfer : Prototype, experiment and validation,'' \emph{IEEE
  Transactions on Wireless Communications}, pp. 1--1, 2020.

\bibitem{bruno_Communications_and_Signals_Design_for_WPT}
Y.~{Zeng}, B.~{Clerckx}, and R.~{Zhang}, ``Communications and signals design
  for wireless power transmission,'' \emph{IEEE Transactions on
  Communications}, vol.~65, no.~5, pp. 2264--2290, May 2017.

\bibitem{Clerckx_rectenna}
J.~{Kim}, B.~{Clerckx}, and P.~D. {Mitcheson}, ``Prototyping and
  experimentation of a closed-loop wireless power transmission with channel
  acquisition and waveform optimization,'' in \emph{Proc. IEEE Wireless Power
  Transfer Conference (WPTC)}, May 2017.

\bibitem{Rectenna_collado}
A.~{Georgiadis}, G.~{Vera Andia}, and A.~{Collado}, ``Rectenna design and
  optimization using reciprocity theory and harmonic balance analysis for
  electromagnetic ({EM}) energy harvesting,'' \emph{IEEE Antennas and Wireless
  Propagation Letters}, vol.~9, pp. 444--446, May 2010.

\bibitem{Adib_Spatial}
\BIBentryALTinterwordspacing
Y.~Ma, Z.~Luo, C.~Steiger, G.~Traverso, and F.~Adib, ``Enabling deep-tissue
  networking for miniature medical devices,'' in \emph{Proc. Conference of the
  ACM Special Interest Group on Data Communication (SIGCOMM)}, Aug. 2018.
  [Online]. Available: \url{http://doi.acm.org/10.1145/3230543.3230566}
\BIBentrySTDinterwordspacing

\bibitem{Carvalho_spatial_combining}
A.~J.~S. {Boaventura}, A.~{Collado}, A.~{Georgiadis}, and N.~B. {Carvalho},
  ``Spatial power combining of multi-sine signals for wireless power
  transmission applications,'' \emph{IEEE Transactions on Microwave Theory and
  Techniques}, vol.~62, no.~4, pp. 1022--1030, Apr. 2014.

\bibitem{GKKurt}
E.~Davut, O.~Kazanci, A.~Caglar, D.~Altinel, M.~B. Yelten, and G.~K. Kurt, ``A
  test-bed based guideline for multi-source energy harvesting,'' in \emph{Proc.
  10th International Conference on Electrical and Electronics Engineering},
  Nov. 2017.

\bibitem{Kurt_Effects_of_Different_Modulation_Techniques_on_Charging_Time_in_RF_Energy_Harvesting_System}
M.~{Cansiz}, D.~{Altinel}, and G.~K. {Kurt}, ``Effects of different modulation
  techniques on charging time in {RF} energy-harvesting system,'' \emph{IEEE
  Transactions on Instrumentation and Measurement}, vol.~69, no.~9, pp.
  6904--6911, Sep. 2020.

\bibitem{Tsolovos}
S.~Nikoletseas, T.~P. Raptis, A.~Souroulagkas, and D.~Tsolovos, ``Wireless
  power transfer protocols in sensor networks: Experiments and simulations,''
  \emph{Journal of Sensor and Actuator Networks}, vol.~6, no.~2, Apr. 2017.

\bibitem{SJha1}
K.~Li, C.~Yuen, and S.~Jha, ``Fair scheduling for energy harvesting {WSN} in
  smart city,'' in \emph{Proc. 13th ACM Conference on Embedded Networked Sensor
  Systems}, Nov. 2015.

\bibitem{GChen}
H.~Dai, X.~Wang, A.~X. Liu, F.~Zhang, Y.~Zhao, and G.~Chen, ``Omnidirectional
  chargability with directional antennas,'' in \emph{Proc. 24th IEEE
  International Conference on Network Protocols}, Nov. 2016.

\bibitem{Rabaey}
P.~S. Yedavalli, T.~Riihonen, X.~Wang, and J.~M. Rabaey, ``Far-field {RF}
  wireless power transfer with blind adaptive beamforming for {Internet of
  Things} devices,'' \emph{IEEE Access}, vol.~5, pp. 1743--1752, Feb. 2017.

\bibitem{RHoward}
X.~Fan, H.~Ding, S.~Li, M.~Sanzari, Y.~Zhang, W.~Trappe, Z.~Han, and R.~E.
  Howard, ``Energy-ball: Wireless power transfer for batteryless {Internet of
  Things} through distributed beamforming,'' \emph{Proc. ACM on Interactive,
  Mobile, Wearable and Ubiquitous Technologies}, vol.~2, no.~2, Jul. 2018.

\bibitem{SWIPT_Clerckx}
J.~{Kim}, B.~{Clerckx}, and P.~D. {Mitcheson}, ``Experimental analysis of
  harvested energy and throughput trade-off in a realistic {SWIPT} system,'' in
  \emph{Proc. IEEE Wireless Power Transfer Conference (WPTC)}, Jun. 2019.

\bibitem{SWIPT_Claessens}
S.~{Claessens}, N.~{Pan}, D.~{Schreurs}, and S.~{Pollin}, ``Multitone {FSK}
  modulation for {SWIPT},'' \emph{IEEE Transactions on Microwave Theory and
  Techniques}, vol.~67, no.~5, pp. 1665--1674, May 2019.

\bibitem{datasheet_P2110B}
{Powercast}. P2110b module datasheet. [Online]. Available:
  “https://www.powercastco.com/documentation/p2110b-module-datasheet/",
  Accessed on 27 Aug. 2020.

\bibitem{Bruno_DC-DC_Optimal_Operation_of_Multitone_Waveforms_in_Low_RF-Power_Receivers}
M.~H. {Ouda}, P.~{Mitcheson}, and B.~{Clerckx}, ``Optimal operation of
  multitone waveforms in low {RF}-power receivers,'' in \emph{Proc. IEEE
  Wireless Power Transfer Conference (WPTC)}, Jun. 2018.

\bibitem{Bruno_contrast_Robust_Wireless_Power_Receiver_for_Multi-Tone_Waveforms}
M.~H. {Ouda}, P.~D. {Mitcheson}, and B.~{Clerckx}, ``Robust wireless power
  receiver for multi-tone waveforms,'' in \emph{Proc. 49th European Microwave
  Conference (EuMC)}, Oct. 2019.

\bibitem{Yoo_A_Dual-Circular-Polarized_Endoscopic_Antenna_With_Wideband_Characteristics_and_Wireless_Biotelemetric_Link_Characterization}
A.~{Basir}, M.~{Zada}, Y.~{Cho}, and H.~{Yoo}, ``A dual-circular-polarized
  endoscopic antenna with wideband characteristics and wireless biotelemetric
  link characterization,'' \emph{IEEE Transactions on Antennas and
  Propagation}, pp. 1--1, Jun. 2020.

\bibitem{Yoo_A_Metamaterial-Coupled_Wireless_Power_Transfer_System_Based_on_Cubic_High-Dielectric_Resonators}
R.~{Das}, A.~{Basir}, and H.~{Yoo}, ``A metamaterial-coupled wireless power
  transfer system based on cubic high-dielectric resonators,'' \emph{IEEE
  Transactions on Industrial Electronics}, vol.~66, no.~9, pp. 7397--7406, Sep.
  2019.

\end{thebibliography}

\vspace{-30pt}
\begin{IEEEbiography}[{\includegraphics[width=1in,height=1.25in,clip,keepaspectratio]{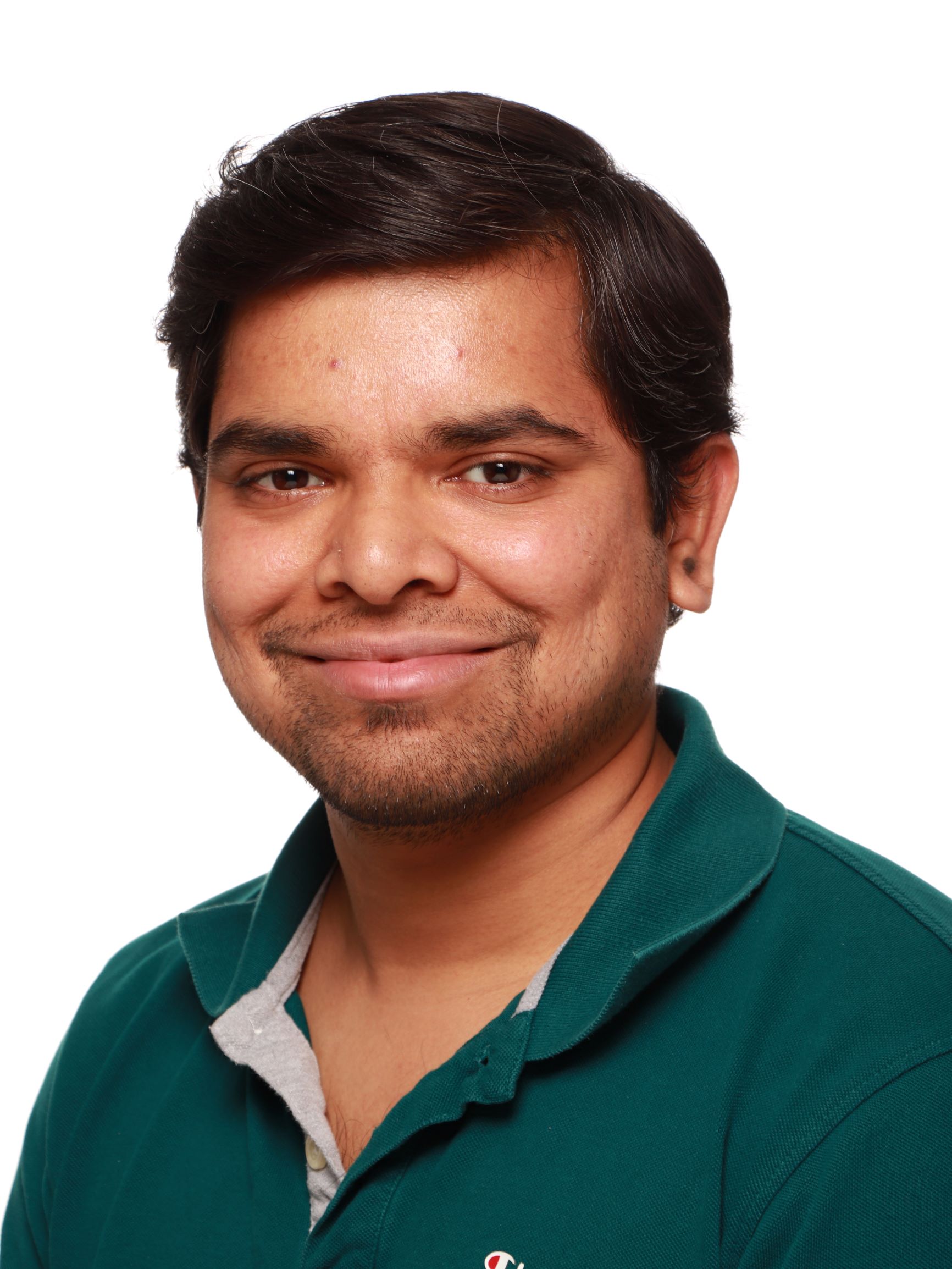}}]{Nachiket Ayir} received his M.Sc. by Research degree in electronics and communications engineering from International Institute of Information Technology, Hyderabad, India in May 2018. He is currently with the Faculty of Information Technology and Communication Sciences at Tampere University as a Doctoral researcher. His current research interests are in the field of wireless energy harvesting, optimization techniques, software-defined radios, simultaneous wireless information and power transfer.
\end{IEEEbiography}

\vspace{-30pt}
\begin{IEEEbiography}[{\includegraphics[width=1in,height=1.25in,clip,keepaspectratio]{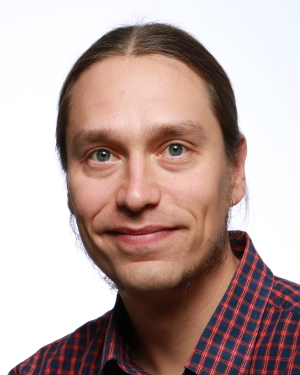}}]{Taneli Riihonen}(S'06--M'14)
received the D.Sc.\ degree in electrical engineering (with distinction) from Aalto University, Helsinki, Finland, in August 2014. He is currently an Assistant Professor (tenure track) at the Faculty of Information Technology and Communication Sciences, Tampere University, Finland. He held various research positions at Aalto University School of Electrical Engineering from September 2005 through December 2017. He was a Visiting Associate Research Scientist and an Adjunct Assistant Professor at Columbia University in the City of New York, USA, from November 2014 through December 2015. He has been nominated eleven times as an Exemplary/Top Reviewer of various IEEE journals and is serving as an Editor for \textsc{IEEE Wireless Communications Letters} since May 2017. He has previously served as an Editor for \textsc{IEEE Communications Letters} from October 2014 through January 2019. He received the Finnish technical sector's award for the best doctoral dissertation of the year and the EURASIP Best PhD Thesis Award 2017. His research activity is focused on physical-layer OFDM(A), multiantenna, relaying and full-duplex wireless techniques with current interest in the evolution of beyond 5G systems.
\end{IEEEbiography}

\vspace{-30pt}
\begin{IEEEbiography}[{\includegraphics[width=1in,height=1.25in,clip,keepaspectratio]{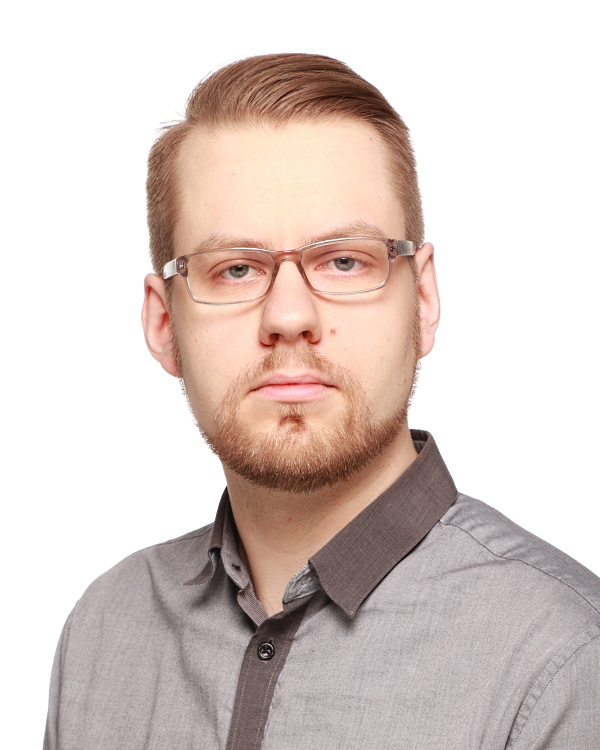}}]{Markus {All\'en}}
was born in Yp\"{a}j\"{a}, Finland, on October 28, 1985. He received the B.Sc., M.Sc. and D.Sc. degrees in communications engineering from Tampere University of Technology, Finland, in 2008, 2010 and 2015, respectively. He is currently with the Faculty of Information Technology and Communication Sciences at Tampere University as a University Instructor. His current research interests include software-defined radios, 5G-related RF measurements and digital signal processing for radio transceiver linearization.
\end{IEEEbiography}

\vspace{-30pt}
\begin{IEEEbiography}[{\includegraphics[width=1in,height=1.25in,clip,keepaspectratio]{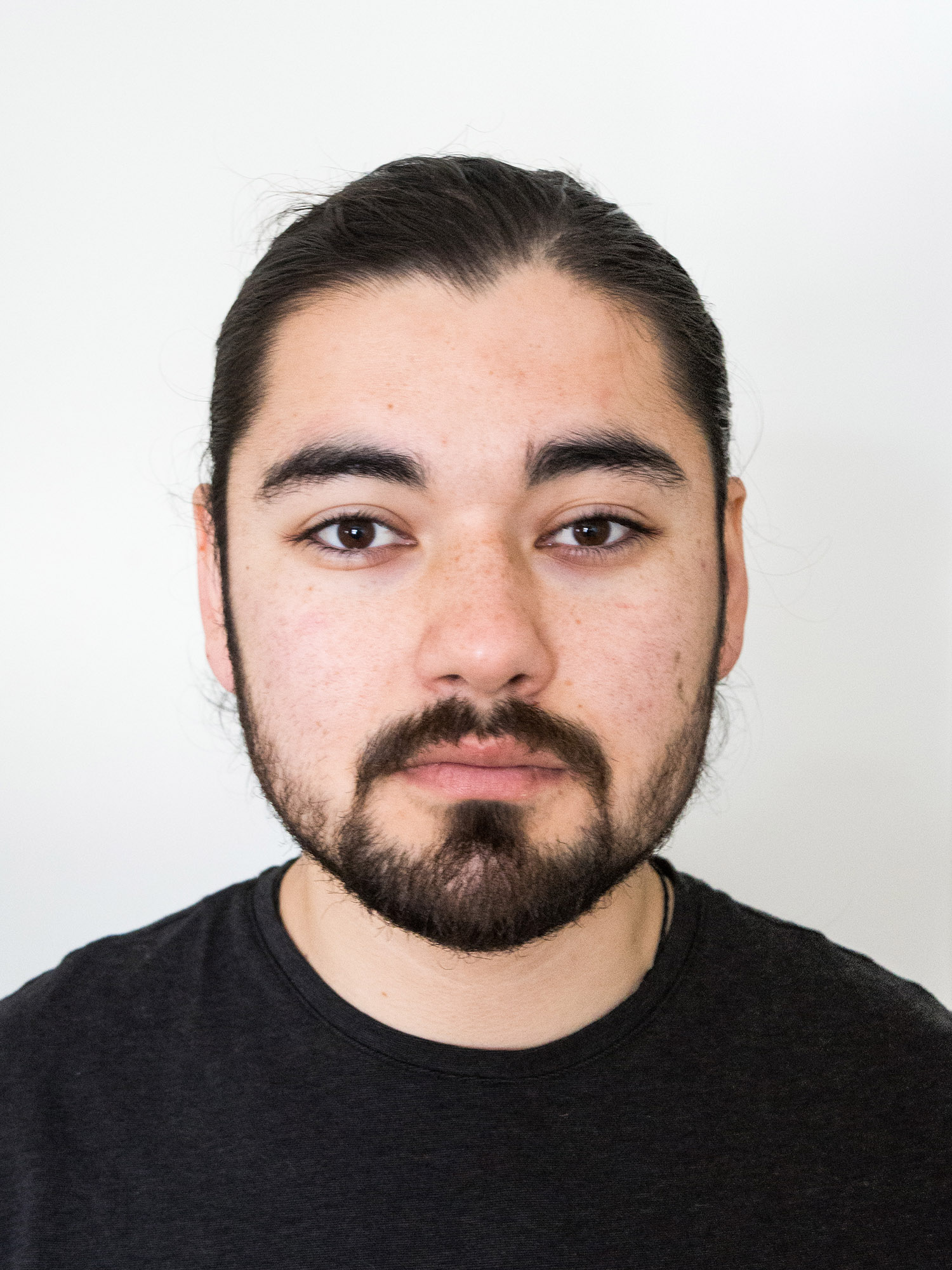}}]{Marcelo Fabi{\'a}n Trujillo Fierro}
achieved the Bachelor's degree in electronics at Universidad Mayor in Santiago de Chile, and the M. Sc. in electrical engineering from Tampere University of Technology. From May until November 2018 he worked in the department of Electronics and Communications Engineering, Tampere University of Technology, Finland. His Master degree's thesis was based in the designing and developing of a research test-bed in the wireless power transfer area by using software defined radios. His areas of interest are a wide scope of different wireless communications technologies such as, cellular networks, IoT, 5G and WPT. 
\end{IEEEbiography}
%\vfill
\end{document}